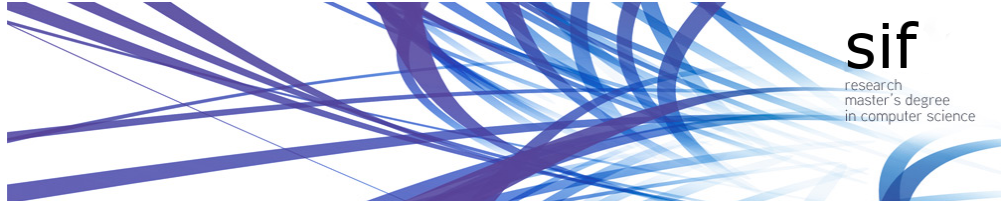

Master research Internship

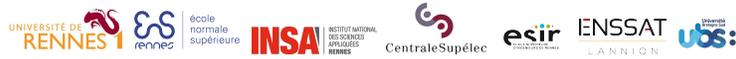

Internship report

# Multi-Channel Automatic Music Transcription Using Tensor Algebra

**Domain: Information Retrieval - Sound**

*Author:*
Axel Marmoret

*Supervisor:*
Nancy Bertin
Jeremy Cohen
PANAMA

**Abstract:** Music is an art, perceived in unique ways by every listener, coming from acoustic signals. In the meantime, standards as musical scores exist to describe it. Even if humans can make this transcription, it is costly in terms of time and efforts, even more with the explosion of information consecutively to the rise of the Internet. In that sense, researches are driven in the direction of Automatic Music Transcription. While this task is considered solved in the case of single notes, it is still open when notes superpose themselves, forming chords. This report aims at developing some of the existing techniques towards Music Transcription, particularly matrix factorization, and introducing the concept of multi-channel automatic music transcription. This concept will be explored with mathematical objects called tensors.

# Contents

# 1 Introduction

Nowadays in western cultures, music is one of the preponderant artistic and human form of expression, as everyone is exposed to it on a daily basis. For this reason, automatic systems need to be able to analyze and help humans in their interactions with it. In particular, musicians wanting to interpret some excerpt of music or to keep trace of an improvisation need a tool to automatically interpret the sounds and rewrite these sounds in an understandable language for them; this is the task of musical transcription. While humans have developed standard languages and have been proven good in transcribing music, the explosion of information nowadays makes humans, at most, inefficient for this task. In addition, they need to be experts in the domain, and are prone to errors. A research field is dedicated to the study of music signals: the Music Information Retrieval (MIR). More specifically, Automatic Music Transcription (AMT, called Music Transcription in this report), the research field for the musical transcription task, is a part of it.

The goal of Music Transcription is far for being achieved yet. Indeed, while musical representation are well defined (with musical scores or tablatures for example), the relations between these representation and the physical signals are hard to represent. In that sense, the first part of this report is dedicated to describing the appropriate representation of Music for its study. In a second part, this report will present the state of the art techniques in Music Transcription. In the following parts, this report will present the work conducted in my internship, firstly by developing the transcription framework and the way of evaluating it, and secondly by developing respectively mono-channel and multi-channel music factorization models. This models will finally be compared on the transcription task on a music database called MAPS.

# 2 Music transcription

## 2.1 Problem formulation

Music Transcription aims at converting the audio signal of an excerpt in a form of musical notation. Hence, the physical characteristics of the signal are at the basis of this task, and need to be defined.

### 2.1.1 Musical notes

As well explained in [1], the signal is generally expressed (at least in western music) in musical notes. Notes are defined according to four parameters:

- The pitch, correlated to the signal frequency (or frequencies). Each note is defined by a fundamental frequency $f_0$, and by harmonics (or "partials"). A note played on different instruments will have the same $f_0$. Partials are generally multiple of this $f_0$, but can be slightly shifted, as presented in Fig. 5b. In this situation, the note and the instrument are called "inharmonic".

- The duration of the note, as each signal is limited in time. As so, we define the onset time, which represent the start of a note, and the offset time, when the signal is weak enough for the note to be considered finished. The offset time lacks clear definition and is hence less critical than onset.

- The loudness, correlated to the amplitude of the signal.

- The timbre, characterized by the temporal and spectral shapes of the signal. Hence, timbre is very large and ill-defined. In fact, timbre represent all that we could not model by the other parameters, and can be seen as the difference between two instruments playing the same note.

As an example, Fig. 1 represents the same note, $A4$, (which is the standard in tuning systems, $la_4$ in the French system, with a fundamental frequency of 440Hz), played by different instruments. The left part shows the acoustic signals of these instruments, and the right shows the frequency spectra of these signals. Peaks on each spectrum represent the fundamental frequency, 440Hz, and its harmonics, multiples of 440Hz. The spectral shapes represent timbres.

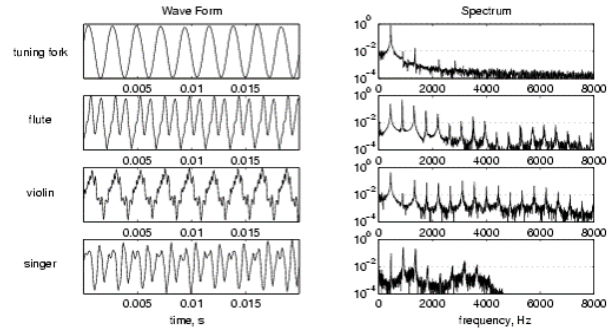

Figure 1: The same note, $A4$, played by different instruments. (http://amath.colorado.edu/pub/matlab/music/)

The goal of Music Transcription is to transcribe a musical signal in a human-readable form. Hence, the key characteristics for transcription are pitch, duration (onset and offset) and amplitude. In my internship, I will focus on pitch and onset/offset.

### 2.1.2 MIDI

MIDI is the desired transcription format. This format represents the frequencies on an integer scale with a step of one semi-tone, and the onset and duration of the note. For example, the $A_4$, with a fundamental frequency of 440Hz, is modeled by the integer 69. It can also represent the amplitude of the note, but is restricted here to fundamental frequency and onset/offset time. Each octave in music represents a 6 tones gap, and so each octave in MIDI is a separation of 12 units. A complete frequency-MIDI conversion table is presented in Fig. 2, along with the standard piano keys.

Hence, the goal of transcription in this internship is to translate musical signals into MIDI files, composed of the pitches and of the onset/duration characteristics. Particularly, I will focus on piano transcription, as the largest database available, MAPS, detailed in Sec. 2.3.1, is composed of piano excerpts. To achieve this goal, the signal needs to be transformed.

Figure 2: MIDI-Pitch Correlation (commons.wikimedia.org/wiki/File:Midi.png)

## 2.2 Signal representation for transcription

As said in Sec. 2.1.1, the scope of this internship is restricted to estimating the pitch and duration of notes. As pitch is defined in the frequency domain, musical signals may be expressed in a time-frequency domain.

The most frequent representation is based on Short-Time Fourier Transformation (STFT) of the signal. This transformation is specifically designed to find frequencies of time varying signals by applying Fourier transformation on small frames of equal length. As a result, we obtain the frequencies of the signal over time. In STFT, frequencies are discretized in uniform bandwidth. However, both time and frequency discretization induce a time-frequency uncertainty principle, which leads to a compromise in the choice of bandwidth [1]. In music nevertheless, the lowest pitched notes, corresponding to the lowest frequencies, are rarely played (first piano keys for example), and time interval between two consecutive notes is in general big enough to address the problem.

The STFT of a signal $x$ is given by the formula:

$$STFT(x)(k,r) = \sum_{n=0}^{N-1} x(rN+n)w(n)e^{-i2\pi kr/N} \quad (1)$$

$k$ and $r$ being respectively the frequency band and time index, $w$ being the time window of a frame and $N$ the size of the frame in discrete time. It computes a $F \times T$ matrix, $F$ being the number of frequency bands for the STFT and $T$ the number of frames.

A time-frequency spectrogram is then computed as the magnitude or the squared magnitude (power spectrogram) of the nonnegative frequencies of the STFT of the signal. A spectrogram is represented in Fig. 3.

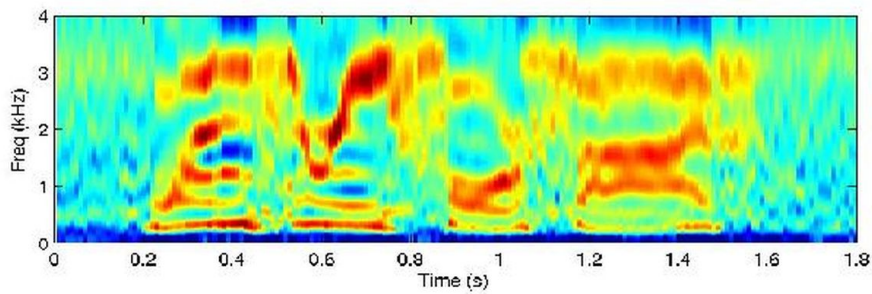

Figure 3: A time-frequency spectrogram (http://info.ee.surrey.ac.uk/Teaching/Courses/eem.ssr/lab2.html)

Rather than being redeveloped, the STFT, in my internship, will be computed using the Scipy Signal Toolbox [2].

Another time discretization technique is to apply a non linear frequency discretization [1]. This is an interesting idea because the fundamental frequencies scale isn't linear: one semi-tone gap is obtained by multiplying the fundamental frequency by $2^{1/12}$. This is the direct consequence of notes construction: the octave of a note is obtained by doubling its fundamental frequency, while every octave contains 6 tones. In that sense are defined Equivalent Rectangular Bandwidth (ERB), which corresponds in frequency bandwidth that contains the same amount of notes. The relation between a frequency in Hertz and in ERB is non trivial: $f_{ERB} = 9.26 \log(0.00437 f_{Hz} + 1)$ (equation IX.2 from [1]). The experiences conducted under that approach concluded that this discretization does not give significantly better results than an uniform discretization, but sensibly decreases computational time. Hence, this technique will not be implemented in my internship.

## 2.3 Data

One of the most important aspect of this work will be to evaluate its transcription performance. To study, develop and compare it, it is necessary to have access to data, in particular large datasets of relevant data. In Music Transcription, this data need to be sound excerpts of notes or music, annotated with the MIDI pitch. Particularly, my internship contribution focuses on multi channel data, which corresponds to musical excerpt containing different tracks. Data containing exactly two channels are called "stereo". Such annotated and stereo musical datasets exist today, and are presented below.

### 2.3.1 MAPS

In piano automatic transcription, the most used dataset is MAPS [3]. MAPS stands for MIDI Aligned Piano Sounds, and contains isolated notes, random pitch and usual western chords (i.e. usual chords in western music), and pieces of music played on pianos. The greatest strength of MAPS is that it has been specifically developed for MIR, and so answers the needs of the community.

Indeed, for systems to be compared equitably, large quantities of annotated data are required. In MIR, this implies large quantities of excerpts to be publicly shared and studied. As any art, music is subjected to copyright laws, complicating this spread of information. Still, as music can be created by anyone knowing how to play an instrument, free-of-right musics exist, but they require time and efforts by humans. In addition, music has to be annotated. In Music Transcription, annotations have the advantage of being objective, as musical notes are clearly defined and standards exist. The direct drawback is that annotator needs to perfectly know these standards, and to easily recognize them. This requires the annotator to have a strong background in music theory. To solve these problems, the excerpts were created from synthetic generator or by player pianos, pianos which play automatically, directly from MIDI files. This results in high-quality music dedicated to this task, with reliable annotations. MAPS contains 40GB of music, which represents about 65 hours of recordings.

These advantages made of MAPS one of the standards in Music Transcription. Indeed, most of the papers today evaluate their results over the MAPS dataset (in this literature review for example, non exhaustively: [1][4][5][6][7][8][9], and is cited in [10]).

The main drawback of MAPS is that it is only composed of piano excerpts, and doesn't take into account the problem of different instruments. Also, as we will see in Sec. 6.2, the stereo information is a complex model of audio mixture.

### 2.3.2 RWC

Another important dataset is the RWC (Real World Computing) music database [11], free of use for researchers. This database is composed of music of three different musical genres: jazz, classical, and J-pop ("popular Japanese music"). These musics were specifically played and recorded for the purpose of the database, with free-of-right musical pieces, professional musicians and sound technicians, and professional recording equipment.

The database contains 215 musical pieces, all annotated with MIDI files.

This database was intended to be a good supplement if the internship realization was proven efficient enough on MAPS, or for testing methods on other instruments than piano, but hasn't been used in practice.

### 2.3.3 Simulated data

Additionally, it could have been useful to make up my own data, to evaluate the robustness of the system on new and different data, or to produce data responding to specific needs. For example, evaluate the impact of more than 2 channels over the system (or even check if it is applicable), or samples with different audio mixtures, could have extend or validate the results. It is one of the perspectives of this work.

## 3 State of the art in single channel transcription

### 3.1 Single, multi-pitch estimation and transcription

Pitch estimation consists in estimating the present notes in an excerpt, and so, finding the present fundamental frequency (or frequencies). Single pitch estimation is the task when only one fundamental frequency is involved, and is considered a solved problem nowadays [10]. However, systems have more difficulties to extend these results at multi-pitch scales (polyphonic signals), when several notes are played at the same time (whether in chords or by different instruments), as musical notes overlap in frequencies and time domains, confusing them. Indeed, it is hard for systems to differentiate partials from fundamental frequencies, and noise from notes. Hence, multi-pitch estimation is the topical subject in Music Transcription and pitch estimation.

More precisely, the Music Transcription problem can be studied through two lenses:

- Framewise estimation of the pitches, which is often called "Multi pitch estimation". This task aims at evaluate the pitches on each frame, and then, aggregate the results to form a global score.

- Notewise estimation of the score. This task aims at detecting notes entirely, with their onset and offset, and to aggregate all the found notes to form a score.

My internship will consist in transcribing a music as a notewise estimation of the score. In that sense, I will only study this case as state of the art, even though some models can treat both cases. Statistical outputs associated to notewise estimation are explained in Sec. 4.3.

### 3.2 Overview of techniques

#### 3.2.1 Frequent techniques

Music Transcription is a long studied task, as it was introduced [12] in 1975 by Moorer [13]. In that sense, lots of transcription models have been developed and studied. In this overview, we will present some of the most frequent techniques.

Statistical models have been largely studied in the context of Machine Learning, as Maximum Likelihood/Maximum A Posteriori and Bayesian classifiers, for example, which have been adapted to the music transcription task [4][14][15][16]. These systems, via statistical estimators for musical elements, as the frequency distribution, loudness, noise, etc, can efficiently model and estimate the musical information and evaluate the probabilities of presence of notes to compute transcription.

Another statistical model adapted to transcription is the Hidden Markov Model (HMM) framework [17][18], which uses a time representation of an isolated note according to sequential states ("Attack, Sustain, Noise" typically [17]) to detect note events.

Specific designed models also use musical high-level features, like harmonicity, smoothness of the sequence or mean bandwidth to compute probable note sequences [19][20]. These features can also be used as estimators in a statistical model [17].

Finally, another frequent technique is based on spectrogram factorization. The spectrogram is represented by a frequency/frame matrix, which is then factorized in low rank non negative matrix, two typically, representing respectively the spectral "bases" for the notes and the activations of these bases over time. This is called NMF for Nonnegative Matrix Factorization [21], and represent the baseline for my internship [1][22]. It is detailed in Sec. 3.3.2.

Additionally, "peak picking", which means keeping the most salient peaks in the spectrogram, is a frequent technique for enhancing transcription by constraining note shapes, but can be used as a whole transcription system [23]. The principle is to detect the highest peaks, and, assuming that they should be fundamental frequencies, try to find its partials, multiple of its frequency. These peaks are then grouped and gives a probability of the presence of the note.

### 3.2.2 Neural Networks approaches

In addition of these techniques, some music transcription approaches rely on neural networks. Today, neural networks have become inescapable due to their results, and their ability of discovering relevant relations in data representation. Indeed, in less than ten years, neural networks have been proven efficient in numerous tasks, especially in Image Recognition. Even if Music Transcription and Image Recognition are different, both belong to signal processing, and can present similarities. This part aims at presenting some neural network approaches in piano transcription.

An interesting approach is proposed in [24] where every note is modeled by its own network. This results in 76 networks dedicated to the recognition of all piano notes except the lowest octave, which led to poor recognition scores. Two notes at the octave are considered as different notes. They used time delay neural networks for this task, a network model which aims at representing time relations within data. In our context, this means that relations between consecutive frames will be modeled. This is particularly useful in Music Transcription as notes can spread among several frames, and knowing the context of a note is necessary for onset/offset detection. The pitch recognition is based on partials group detection, that is, detection of the fundamental frequencies of the note and its partials. Agglomerating all the results of the different partials detection block ("oscillators", up to ten) will lead to an overall score as output value. A high output value indicates the presence of the note, whereas a low input value indicate its absence. This model results in a highly accurate system, but also with a high octave error.

In [8] is proposed an approach for onset detection based on Bidirectionnal Recurrent Neural Network with Long Short-Term Memory (LSTM) units. A Recurrent Neural Network is a multi-layer neural network where neurons in hidden layers not only are connected to the next layer' ones but also to themselves (recurrent connections). This kind of architecture offers the ability for the network to model temporal contexts, as information persists with time. In practice, information is erased in a few iterations with only recurrent connections. LSTM units were introduced to address this problem, replacing the recurrent connections. Finally, this network is bidirectional, which means that the hidden layers are twice as many as in a regular Recurrent Network, but with the new layers receiving the input values in the reversal temporal order. With that architecture, a temporal context is built around an input value, in both temporal directions (in the past and in the future). For onset detection, context is important, as notes have specific physical characteristics during its expression, such as increase in energy for attack (onset), or special energy envelope on the other phases.

In addition, notes are often correlated to their neighbors, due to harmonicity in the construction of a musical piece. Hence, knowing the closest notes can help in pitch recognition. Thus, this architecture seems to be relevant in our context. Still, the complexity of that system should require a huge computational time and computing power, although it is not expressed in the paper.

They obtain high results in onset detection (about 85% in accuracy scores over MAPS (see Sec. 2.3.1) with a temporal window close to the annotation precision).

Finally, a comparative study of different architectures in pitch detection is proposed in [9]: Deep Neural Networks, Recurrent Neural Networks and Convolutional Neural Network. Based on STFT spectrogram, the Convolutional Neural Network performs the best. Nonetheless, it doesn't reach the state of the art performed with NMF estimation techniques on MAPS database [9]. We cannot conclude anything about the overall performance of neural networks in Music Transcription, but it shows that neural networks aren't abruptly increasing performance yet, as it was the case in Image Recognition.

### 3.2.3 Overview of MIREX contest results

As in many research fields, Music Transcription (and MIR in general) systems are evaluated in contests and conferences. The most significant one in MIR is the MIREX contest [25], held as part of the ISMIR Conference. Researchers of the field can compare their systems in different categories, particularly in multi-pitch evaluation. Through the years, many different techniques drove the research. Among all the different algorithms presented in the MIREX, [10] gives an overview of the preponderant techniques since mid-2000s.

Ten years ago, the best results were obtained by evaluating the most probable note sequence according to high-level features [20], or with Maximum Likelihood/Maximum A Posteriori and Bayesian classifiers [4].

These two techniques obtained about 65% accuracy scores on note recognition in recent MIREX contest, and a F-measure about 0.35 in onset/offset detection (with a 20% margin error in the duration of the note), respectively in 2010 [20] and in 2013 [4].

The NMF technique is less present in MIREX contest. In this bibliography, the only submitted paper on MIREX [26], in 2007, obtained about 54% accuracy score in note detection, and a F-measure of 0.2 in onset/offset detection, and was ranked 2nd. This can be explained by the fact that NMF is mostly used in piano note detection [1][26][5], when MIREX contest perform recognition on several instruments. Indeed, as NMF considers a limited number of fundamental frequencies (as a matrix dimension), it is less adapted for instruments with a continuous possible range of fundamental frequencies (as a violin for example) than for piano, where the possible fundamental frequencies are quantified. Still, as a comparison, [5] obtained a 0.8 F-measure in note detection on MAPS, detailed in Sec. 2.3.1 and, in 2010, in the same contest than [20], a NMF based technique [6] obtained accuracy scores of 45% in note detection.

Neural Networks approaches have also been proposed, like in 2016 [24], with a F-measure score of 35.2%, or in 2017 [27], with only a framewise note estimation system.

All the MIREX results are available online [28].

Considering all these techniques, we will now focus on presenting the NMF framework in Music Transcription, and trials at improving it, notably by constraining it and extending it to multichannel signals.

## 3.3 NMF

### 3.3.1 Mathematical model

NMF stands for Non-negative Matrix Factorization. It was brought to light by Lee and Seung [21]. Since then, many papers tried to adapt NMF in many fields, and Music Transcription in particular. In this model, the variable, a nonnegative matrix $X \in \mathbb{R}_+^{K \times N}$, will be decomposed as:

$$X \approx WH \tag{2}$$

where $W \in \mathbb{R}_+^{K \times R}$ and $H \in \mathbb{R}_+^{R \times N}$, component-wise nonnegative [29]. This technique can be used as a dimensionality reduction technique, implying typically $R \ll K, N$.

Denoting $\|.\|_F$ the Frobenius norm of a matrix, one popular way of finding the best factorization is to solve the following optimization problem:

$$\min_{W \geq 0, H \geq 0} \|X - WH\|_F^2 \tag{3}$$

with $\|X - WH\|_F^2 = \sum_{i,j}(X_{ij} - (WH)_{ij})^2$.

This formulation poses some problems, though:

- This problem is proven NP-Hard for exact decomposition of $X$ [30], and is not convex when considering both matrices [31]. Some heuristics still find results, but may fall on local minima and be highly complex.

- Under that form, this optimization problem may admit many solutions: for any $H$ and $W$ solutions of the problem, $HQ$ and $Q^{-1}W$ can be solution as long as $HQ$ and $Q^{-1}W$ are component-wise nonnegative. In practice, regularization or specific constraints over the solutions can reduce the number of solutions, sparsity constraints notably [32].

- On that objective function, $R$, the maximal rank of the matrices $W$ and $H$, is fixed. Hence, finding an adequate rank must be done either by tests, which increases complexity, or with prior knowledge.

Still, this NMF formulation gives good results in particular cases, as prior knowledge and constraints help reducing the solution space. Particularly, sparsity is a well-studied and relevant constraint [32].

To solve this minimization problem, many algorithms can be implemented [29]. A common idea of many algorithms is to optimize alternatively over only one of the two factors, keeping the other fixed. The reason behind this is that the problem becomes convex in the case of one factor to optimize, for example $\min_{W \geq 0} \|X - WH\|_F^2$. Optimizing the other factor can be made easily since the problem is symmetric in $H$ and $W$: $\|X - WH\|_F^2 = \|X^T - H^T W^T\|_F^2$. In that sense, [29] wrote this standard algorithm for NMF:

---

**Algorithm 1** Principle of most NMF algorithms

---

**Input:** Nonnegative matrix $X \in \mathbb{R}_+^{K \times N}$ and a factorization rank $R$
**Output:** $W \in \mathbb{R}_+^{K \times R}, H \in \mathbb{R}_+^{R \times N}$, such that $X \approx WH$ and $rank(W) = rank(H) = R$
1: Generate $W^{(0)}$ and $H^{(0)}$, initial matrices (see below further details)
2: **for** $t = 1, 2, ...$ **do**
3: $\quad W^{(t)} = update\left(X, H^{(t-1)}, W^{(t-1)}\right)$
4: $\quad H^{(t)^T} = update\left(X^T, W^{(t)^T}, H^{(t-1)^T}\right)$

---

Initialization of the matrices can be at random, or with prior knowledge over the right factorization (with expert rules, or by clustering over the columns of $X$ for example). The initialization depends on the problem and on the researchers intuition.

Analogously, the update rule is left to the discretion of the researcher, as it depends on the constraints and on the factorization goal. One unconstrained and common update rule is the Alternated Nonnegative Least Squares [5][29], which will be used in my internship and developed in Sec. 5.1.1.

In this problem, the metric is the Frobenius norm [29]. NMF can also be defined with the Kullback-Leibler divergence or the Itakura-Saito divergence [33].

### 3.3.2 NMF adapted to Music Transcription

This NMF model can be applied to Music Transcription [1][26][5]. In that context, the $X$ matrix represents the input magnitude (or square magnitude for a power spectrogram) in frequency/time bins (module of the STFT coefficients), which will be decomposed in two matrices of lower rank: $W$ for the spectral information, and $H$ for the temporal information. This technique makes sense when we look at it as a dictionary decomposition method; $W$ representing a codebook of the notes according to their frequencies, and $H$ the activations of these notes with time. This analogy needs to be held carefully as, in this model, $W$ is to be found by the algorithm, and not fixed by the researchers as in the traditional dictionary decomposition techniques.

Component-wise, $X \approx \sum_{r=1}^{R} w_{:r} h_{r:}$ with $w_{:r}, h_{r:}$ respectively the $r$-th column of $W$ and row of $H$. The sense of this decomposition is for $w_{:r}$ to represent the frequency decomposition of the $r$-th note, and for $h_{r:}$ the activation of this note in time. An example of decomposition is given in Fig. 4.

Hence, the key of this decomposition is finding the matrix $W$ that best fits the spectral decomposition of notes. The closest the dictionary is to the real frequency definition of notes, the most accurate the activations will be. Conversely, if $W$ elements are different from musical notes, the activations according to this dictionary won't be relevant, and our transcription meaningless.

On the other hand, we could initialize $W$ as the dictionary of notes fundamental frequencies and then observe their activations according to time. Unfortunately, the timbre of a note (its spectral envelope) differs between instruments. It implies that two different instruments would

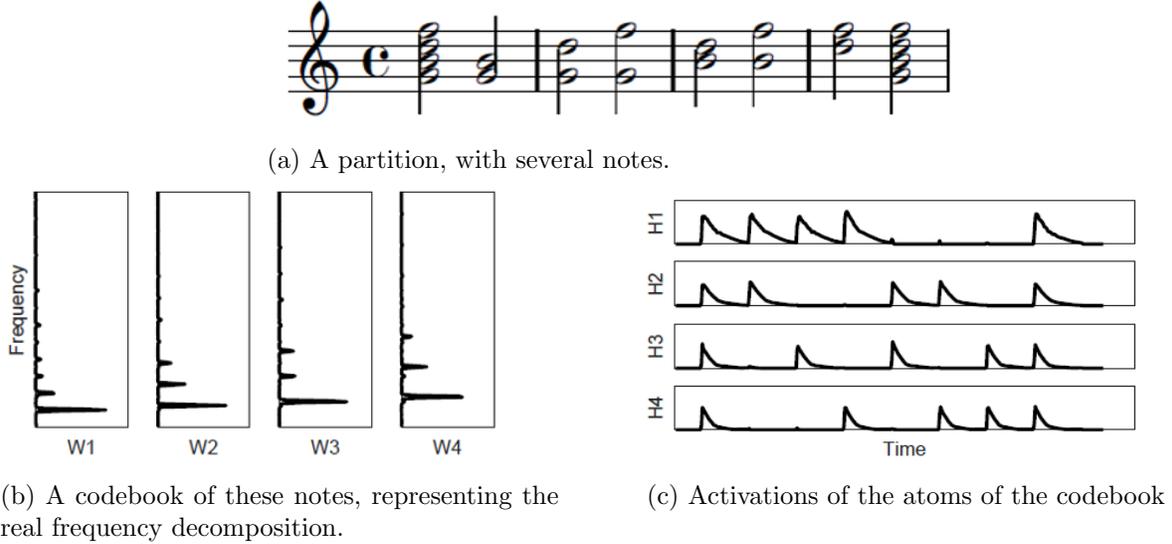

(a) A partition, with several notes.

(b) A codebook of these notes, representing the real frequency decomposition.

(c) Activations of the atoms of the codebook

Figure 4: An example of NMF decomposition (from [1])

need two different dictionaries for their notes to be recognized. In addition, timbre can greatly differ in a family of instruments. Hence, a fixed-dictionary decomposition could potentially overfit the spectral envelopes of the notes of a specific instrument.

In that sense, the perfect way of resolving the transcription task for all instruments should then be between these two extremes: an unconstrained model, which doesn't converge, and a fixed dictionary, which can be too specific for multi-instrument transcription.

Some interesting approaches have been made in that direction.

For example, [26] represents the dictionary as an association of narrowband partials, fixed and not varying according to the instrument playing, and of the spectral envelope, dependant of the instrument playing. The envelope is modeled by weighting the narrowband partials, $W_{if} = \sum E_{ik} P_{ikf}$. This architecture leads to considering harmonicity between partials and their spectral envelopes independently, which should make the system more robust to spectral envelope variations.

Another technique from [5] is based on sparsity decomposition techniques. The main idea is that the most accurate notes lay on several specific frequencies (the fundamental and some partials), and can hereby be considered as sparse frequency vectors. Hence, the goal is to find the sparsest corresponding dictionary. The results presented in the paper shows a good potential for this approach.

More generally, NMF is used with some specific constraints, as harmonicity between elements of the dictionary or relation on differences between consecutive partials for examples. It still is an open problem to find the most relevant and successful constraints.

### 3.3.3 Fundamental frequency estimation in $W$

Finally, the pitch has to be computed from the frequency decomposition in $W$. A natural approach is to find the predominant periodicity between the partials, which should be the common multiplier of the partials, and so, the fundamental frequency [34].

Nonetheless, this technique assumes that the signal is the spectrum of exactly one note, possibly with noise or imprecision, and that the signal is harmonic. Under these assumptions, this periodicity can be found by computing the autocorrelation product of the Inverse Fourier Transform on the spectrum of a note [34]. This is not the only fundamental frequency estimation technique [1], but is the chosen one for my evaluation. This product is maximal at the periodicity points, as the partials, most salient peaks of the signal, are multiplied together. The value at zero isn't considered as the signal multiplied by himself does not give information. As the fundamental frequency is the most salient peak of all, the maximum of this product is the fundamental frequency.

The quality of this estimation can be computed by the pitch salience [35], relation between the two highest peaks, notably to check whether the signal is periodic. Indeed, a low pitch salience indicates that the delayed versions of the signal can't align on partials, and so that any frequency has enough harmonics to be considered as the fundamental frequency of the signal. In our case, this relation is used to check whether the signal is periodic enough to be a note. However, this relation does not inform about the accuracy of the found fundamental frequency.

With that technique, decomposition in $W$ have to be as precise as possible. Indeed, in addition to computing accurate activations in $H$, precise note-vectors enhance $f_0$ estimation, as it relies on correct periodicity. In that sense, missing a few partials can mislead this estimation.

In addition, in the case of piano, the signal is generally inharmonic for the highest notes [36], due to the physical characteristics of the piano. However, in MAPS, most of played notes are the middle ones (see a distribution of the notes in MAPS in Fig. 5a below). In Fig. 5b, the inharmonicity relation is expressed in cents, 100 cents corresponding to a semi-tone shift. As the midi scale is an integer scale, the frequency-midi conversion formula involves a rounding function:

$$f_{MIDI} = round\left(69 + 12 \times \log_2\left(\frac{f_{Hz}}{440}\right)\right) \tag{4}$$

In that sense, small shifts (smaller than 50 cents typically when partials are periodic) shouldn't affect the fundamental frequency estimation, as long as inharmonicity is periodic. This is consistent with findings in [37], when integrating a fixed inharmonicity does not improve results. Still, learning and integrating the specific inharmonicity of the piano to transcribe could enhance results.

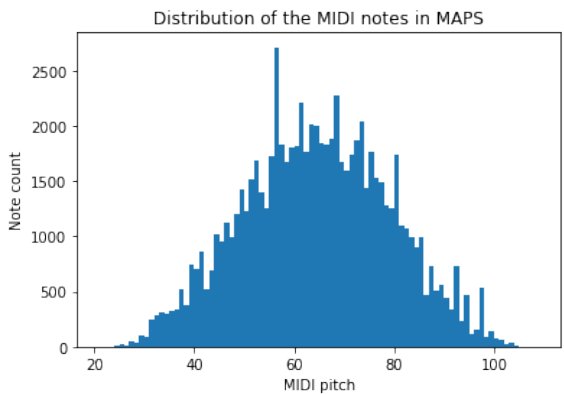

(a) Histogram of distribution of midi notes in MAPS (computed from ground truth)

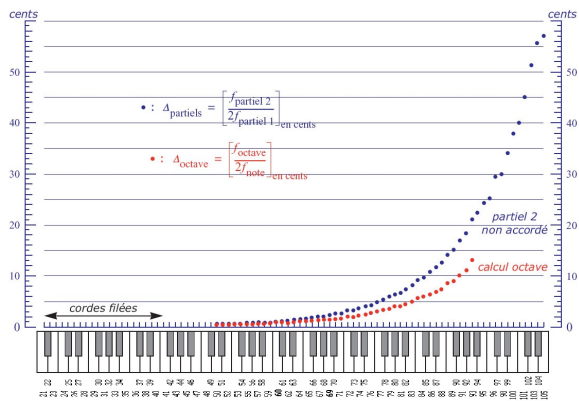

(b) Inharmonicity of the piano (From Claude Valette — Public Domain, https://commons.wikimedia.org/w/index.php?curid=3458867)

Figure 5: Distribution of notes in MAPS compared to the inharmonicity

Finally, rather than learning specific inharmonicity of the piano, and without reliable estimation of the $f_0$, we decided to entirely learn note's spectra in a second part of the internship (see Sec. 5.1.3).

The final stake of my internship being the music transcription, an important step is the transformation from the factorization results in two matrices $W$ and $H$ to a MIDI file, and comparing it the ground truth. This is the goal of the next section.

# 4 Transcription evaluation

## 4.1 Detecting a note

The first goal of transcription is to detect notes, and also not to detect what aren't notes. This is performed by applying a detection threshold on the activation of a note, which is a common technique [1]. The starting idea is that the activation has to be greater than a certain value for considering that a note is being played, and not noise. When an activation is greater that the threshold in $H$, the onset is given by its column, and the row index indirectly indicate the pitch. Indeed, the $i$-th row of $H$ corresponds to the $i$-th column of $W$, which is a note.

More precisely, the ground truth in MIDI indicates the moment where the instruction of playing a note is given, and not the peak of amplitude of the note. Recordings being made on real player pianos, a little delay has to be considered between the ground truth onset and the maximum of the peak. In that sense, taking the column index as onset isn't enough. Instead, starting from the column index where activation is higher than the threshold, onset is given as the moment where the activation is higher than 10% of the threshold if there is such an index in the five previous frames, or is shifted by five frames backwards otherwise. This computation of onset time enhanced the results, and is presented on Fig. 6.

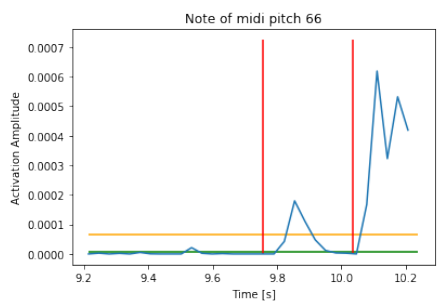

Figure 6: A visualization of the threshold. The vertical red lines represent the onset ground truth, the blue curve represents the activation over time, the orange horizontal line represent the value of the threshold, and the green line represent 10% of the threshold, the threshold used to find a more accurate onset.

Two types of thresholding have been tested, called "Fixed" and "Adaptive".

The "Fixed" threshold is a technique where the threshold value $thresh$ is constant. For a note to be detected, the average of its activation values on five consecutive frames (the current time index and the 4 following) have to be greater than $thresh$. Every frame is 64ms long, and overlaps the previous by half. Averaging the activation on five consecutive frames intends to reduce spurious noise peaks misdetection and several detection of the same note when activations are close to the threshold.

The "Adaptive" threshold technique is inspired from [38]. The idea is to compute a contextual threshold, more robust to stationary noise. To that extent, activations of a note are averaged on 21 consecutive frames, from $t-10$ to $t+10$ for any given $t$ (extremes are padded with zeros), and added to a fixed value. This results in a threshold influenced by the local variations of music, increasing when activations are high. Still, a note attack activation should be detected, as it represents a very localized peak. A visualisation of this thresholding technique is presented in Fig. 7.

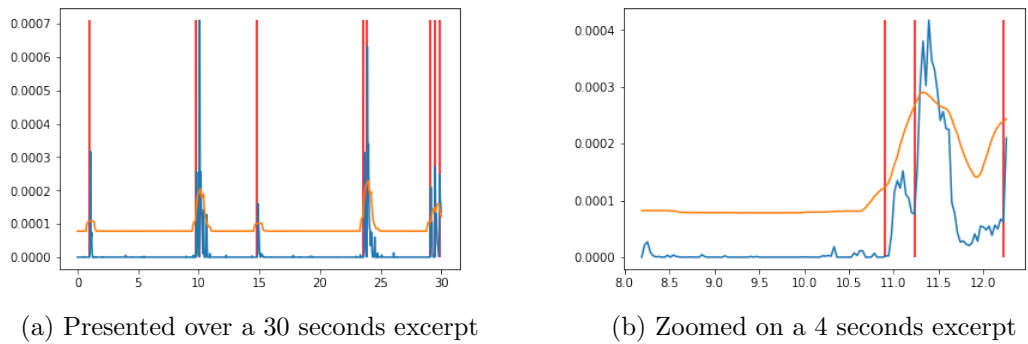

(a) Presented over a 30 seconds excerpt       (b) Zoomed on a 4 seconds excerpt

Figure 7: An example of adaptive threshold on activations, with the same legend than in Fig. 6.

In both techniques, thresholds are computed with a constant part: the threshold itself in the "fixed" technique, the basis added to the consecutive frame average in the "adaptive". This fixed part is computed as a decibel reduction of the maximum of $H$, $\delta_{dB} = 10\log_{10}\frac{\max(H)}{thresh}$, as in [38]. This decibel reduction differs between tests, and strongly impacts the transcription results. In that sense, the results will be presented as the best ones according to the decibel reduction level.

The offset time is set at the moment when activation fall again below the threshold.

### 4.2 Midi writing

Once the notes are detected, MIDI files can be written. This was made with the module MIDIUtil [39], from the onset/offset times and the pitch of each note.

### 4.3 Statistical outputs

To concretely evaluate the quality of our transcription, statistical outputs need to be calculated. Some python libraries calculate these outputs, like mir_eval [40], but I decided to develop my own metrics module in order to understand and learn the way of calculate them. This module calculates the True Positives (notes correctly detected), False Positives (notes detected but which don't exist) and False Negatives (notes in the ground truth that were not transcribed). A note was declared correctly detected when the MIDI pitch was the good one and when the onset time was equal to the ground truth one with a little tolerance. This tolerance is typically 50ms in MIREX [25] or in results in general [38], but was set to 150ms in my final results. This choice was made because the heart of my internship is to compare the factorization potential of each method. From that perspective, we wanted to diminish as much as possible the degradation in the results induced by a delayed detected onset time, as this is not caused by the factorizing technique but rather by the note detection technique. This way, we hoped to stretch as much as possible gaps between the different models.

No distinction were made between pitch errors (a good onset but a wrong pitch) and onset errors (a correct pitch but a wrong onset). The offset time is not taken in account in these outputs, because we don't consider it as well-defined enough.

Results obtained by this implementation were compared and found equals to the ones obtained with mir_eval [40], which validates the correct implementation of this toolbox.

# 5 Implementing NMF and improving it with spectrum learning and sparsity priors

Starting from these theoretical considerations, an important part of my work has been the implementation and tests of NMF techniques. Particularly, we focused on improving NMF with prior knowledge about notes for $W$ and constraints about the activations for $H$.

## 5.1 NMF implementation

### 5.1.1 NMF resolution algorithm

The NMF algorithm was implemented using block-coordinate descent, as in the Alg. 1. More specifically, each block is updated using a Nonnegative Least Squares problem resolution, where both matrices are updated with the same update rule.

As the problem isn't convex (when updating both matrices) [31] and NP hard [30], initialization impacts the output of most NMF algorithms. In that sense, [41] proposed an initialization algorithm called NNDSVD for Non Negative Double Singular Value Decomposition. This algorithm first performs an SVD on the spectrogram $X$, and then performs another SVD on the positive part of each of the resulting SVD factors, using them to initialize $W$ and $H$. The interest of this technique is that the reconstruction error $||X - WH||_F$ is bounded [41](Proposition 6). This theoretical result is extended to practice where their numerical tests suggest a fast decline of the reconstruction error, and an enhancement of the sparsity of the factorization. In our tests, we observe the same results, which are a faster convergence and a decomposition of the codebook $W$ which better approach notewise frequency decomposition than with random initialization.

The NNDSVD initialization algorithm was taken from the NIMFA toolbox [42] rather than being re implemented.

In their seminal paper [21], Lee & Seung defined the following update rule, often denoted as the "Multiplicative Update":

$$W^{(t+1)} = W^{(t)} \times \frac{XH^{(t)T}}{W^{(t)}H^{(t)}H^{(t)T}} \quad (5)$$

with $. \times .$ the component-wise multiplication and $\frac{[.]}{[.]}$ the component-wise division of the matrices. Updating $H$ is analogous when transposing the problem, as in Alg. 1. They proved that this update rule monotonically decreases the Frobenius norm of the approximation $||X - WH||_F$. However, the convergence of this algorithm can be really slow with dense matrices [43].

Hence, we chose instead the HALS update rule, which stands for Hierarchical Alternating Least Squares. This algorithm outperforms the convergence time of the multiplicative update in practice [44]. Instead of updating the entire matrices $W$ and $H$ at each iteration, HALS updates each rank one factor $W_{:k}$ and $H_{k:}$. Rewriting $||X - WH||_F$ as $||X - \sum_{i \neq k} W_{:i}H_{i:} - W_{:k}H_{k:}||_F$, it is possible to define $Res_k$ as the $k$-th residual matrix, $Res_k = X - \sum_{i \neq k} W_{:i}H_{i:}$. The optimisation problem then becomes: $\min_{W_{:k} > 0} ||Res_k - W_{:k}H_{k:}||_F^2$ for every $k$ between 1 and the rank $R$ (and similarly for $H$). The algorithm updates alternatively all columns $W_{:k}$ and all rows $H_{k:}$, as in the original formulation, because both sub problems are convex. These sub problems are resolved as a projected gradient method in [45], in the Lemma 4.4, giving the following closed form update rule for W:

$$W_{:k}^{(t+1)} = \max\left(0, \frac{Res_k^{(t)} H_{k:}^{(t)T}}{||H_{k:}^{(t)}||_2^2}\right) \quad (6)$$

The update rule for $H$, as in Alg. 1, is similar with a transposition of the problem, $H_{k:}^{(t+1)} = \max\left(0, \frac{W_{:k}^{(t+1)T} Res_k^{(t)}}{||W_{:k}^{(t+1)}||_2^2}\right)$. The algorithm obtained with this update rule is proved to converge to stationary points in the Theorem 4.5 [45].

In addition, to gain in computation time, the matrices products are factorized as much as possible, to avoid recalculation in the inner loop. Developing $Res_k$, the update rule for $W_{:k}$, $\frac{Res_k^{(t)} H_{k:}^{(t)T}}{||H_{k:}^{(t)}||_2^2}$, is rewritten as: $\frac{XH_{k:}^{(t)T} - \sum_{i=0}^{k-1} W_{:i}^{(t+1)} H_{i:}^{(t)} H_{k:}^{(t)T} - \sum_{i=k+1}^{R} W_{:i}^{(t)} H_{i:}^{(t)} H_{k:}^{(t)T}}{||H_{k:}^{(t)}||_2^2}$. In that form, we can compute $XH^{(t)}$ and $H^{(t)}H^{(t)T}$ outside of the loop and evaluate respectively the $k$-th column and the ($i$,$k$)-th element, thus reducing the number of matrices product for each column. Finally, doing the same thing with the update rule for $H$ leads to the usual HALS algorithm:

More precisely, we used an accelerated version of HALS [44], where the same matrix ($W$ or $H$) is updated several times in a row (keeping the other fixed), so as to reuse the pre-computed matrices, as these multiplications are the most expensive steps. This results in an inner loop in the updating step. This loop is implemented with two stopping conditions, to avoid losing computation time when the update reaches a minimum or a saddle point. This new algorithm is described in Alg. 3 [44].

The first one consists in estimating the computation time of the first iteration, when matrices $A$ and $B$ have to be pre-computed, compared to the next ones when it won't be the case, and so, the computation time gained when updating the same matrix two times in a row. In practice, a factor $\rho$ is calculated by dividing the computation time of an updating iteration with the calculation of matrices by the computation time without its calculation. Then, the number of iteration will be lower than $\lfloor 1 + \alpha\rho \rfloor$, $\alpha$ being a multiplicative parameter to handle the number of iteration (typically 1/2 in our tests). In other words, this condition consists in spending the same computation time than if the algorithm was performing $\lfloor 1 + \alpha \rfloor$ updates with the calculation of $A$ and $B$. This limits the computation time spent on the update of each factor.

The second condition consists in exiting from the loop when the update is converging, which is guaranteed since the problem is convex when one of the two factors is fixed. This is obtained by computing the update variation at the first iteration of the inner loop, and stopping when the variation in this loop becomes small compared to this value. Mathematically, noting $W^{(t,l)}$ the matrix $W$ at the t-th iteration of HALS and at the l-th iteration of the inner loop, the exiting

**Algorithm 2** HALS algorithm

**Input:** Nonnegative matrix $X \in \mathbb{R}_+^{K \times N}$ and a factorization rank $R$
**Output:** $W \in \mathbb{R}_+^{K \times R}, H \in \mathbb{R}_+^{R \times N}$, such that $X \approx WH$
  Initialize $W^{(0)}$ and $H^{(0)}$ with NNDSVD algorithm.
  **for** $t = 1, 2, ...$ **do**
    Compute $A = XH^{(t)T}$ and $B = H^{(t)}H^{(t)T}$.
    **for** $k = 1, 2, ..., R$ **do**
      **if** $B_{kk} \neq 0$ **then**
$$W_{:k}^{(t+1)} = \max\left(0, \frac{A_{:k} - \sum_{i=0}^{k-1} B_{ik} W_{:i}^{(t+1)} - \sum_{i=k+1}^{R} B_{ik} W_{:i}^{(t)}}{B_{kk}}\right)$$
    Compute $C = W^{(t+1)T} X$ and $D = W^{(t+1)T} W^{(t+1)}$.
    **for** $k = 1, 2, ..., R$ **do**
      **if** $D_{kk} \neq 0$ **then**
$$H_{k:}^{(t+1)} = \max\left(0, \frac{C_{k:} - \sum_{i=0}^{k-1} D_{ki} H_{i:}^{(t+1)} - \sum_{i=k+1}^{R} D_{ki} H_{i:}^{(t)}}{D_{kk}}\right)$$

criterion is: $||W^{(t,l+1)} - W^{(t,l)}||_F \leq \epsilon * ||W^{(t,1)} - W^{(t,0)}||_F$, with $\epsilon$ a positive parameter, small compared to one (0.01 in our tests). This limits the computation time by exiting when the algorithm seems to converge.

This leads to the accelerated HALS algorithm, where one step of the update of $W$ is presented below in the Alg. 3 (similar on $H$).

Conversely to the original paper [44], we allow the columns of $W$ and $H$ to be the zero vector. This is because the NMF rank is estimated in our problem, and so can be overestimated, when the theoretical results assumed that the rank was appropriate.

This algorithm is the basic one we will used for NMF and, later, for NTF.

### 5.1.2 Sparsity constraints

To guide or enforce the factorization, constraints can be added to the model. In particular, sparse NMF, which is the NMF model with sparsity constraints, is a common constrained model [46]. Sparsity has many definitions, the most frequent being "to have as many zeros as possible". Mathematically, this means minimizing the pseudo $l_0$ norm of the considered variable [47].

For music transcription, sparsity can be implemented on the columns of each matrix $W$ and $H$. Sparsity enforces to keep only the most salient entries. For $W$, this means that each note must be formed only by its fundamental frequency and partials, because they are the most redundant frequencies of a note (as noise isn't specific to a note). For $H$, this means limiting the activations to the highest, as a low activation is expected to be noise or residual partial of another note.

Hence, we implemented three different sparse NMF models:

- The first model is based on the $l_0$ definition of sparsity, that is, keeping at most $n$ nonzero values. This is generally called "Hard Thresholding". The different tested values of $n$ were

**Algorithm 3** One step of Accelerated HALS on $W^{(t)}$

---

**Input:** Nonnegative matrix $X \in \mathbb{R}_+^{K \times N}$, $W^{(t)} \in \mathbb{R}_+^{K \times R}, H^{(t)} \in \mathbb{R}_+^{R \times N}$, $\alpha$, $\epsilon$ and a factorization rank $R$

**Output:** $W^{(t+1)}$

    Compute $A = XH^{(t)T}$ and $B = H^{(t)}H^{(t)T}$.
    **for** $k = 1, 2, ..., R$ **do**                                                       ▷ Iteration 0 (same as in the HALS, Alg. 2)
        **if** $B_{kk} \neq 0$ **then**

$$W_{:k}^{(t,1)} = \max \left( 0, \frac{A_{:k} - \sum_{i=0}^{k-1} B_{ik} W_{:i}^{(t,1)} - \sum_{i=k+1}^{R} B_{ik} W_{:i}^{(t,0)}}{B_{kk}} \right)$$

    $\rho = W_{:k}^{(t,1)} - W_{:k}^{(t,0)}$
    **for** $l = 1, 2, ..., \lfloor 1 + \alpha \rho \rfloor$ **do**                                                                ▷ Inner loop
        **for** $k = 1, 2, ..., R$ **do**
            **if** $B_{kk} \neq 0$ **then**

$$W_{:k}^{(t,l+1)} = \max \left( 0, \frac{A_{:k} - \sum_{i=0}^{k-1} B_{ik} W_{:i}^{(t,l+1)} - \sum_{i=k+1}^{R} B_{ik} W_{:i}^{(t,l)}}{B_{kk}} \right)$$

        **if** $||W^{(t,l+1)} - W^{(t,l)}||_F \leq \epsilon * ||W^{(t,1)} - W^{(t,0)}||_F$ **then**
            break.
    Similarly for $H$ and for every t iteration of HALS.

---

    between 10 and 20, because 10 partials for $W$ is empirically enough to compute the fundamental frequency, and because humans are unexpected to play more than 20 notes at the same time (10 at most fingers playing, added to notes potentially sustained with the pedal). When sparsity is enforced on both factors, the same sparsity level is applied. We will denote this technique "$l_0$ Hard Sparse".

- The second model enhanced sparsity by adding a penalty term to the update rule. This penalty term is the $l_1$ norm of the current column multiplied by a constant. Hence, the nonnegative least square to solve for $W_{:k}$ (and analogously for every $k$-th column and for $H$) becomes: $\min_{W_{:k} > 0} ||X - WH||_F^2 + 2\alpha_{sp}||W_{:k}||_1$, with $\alpha_{sp}$ the constant controlling the sparsity level of $W_{:k}$ (the bigger, the sparser). Under that form, the update rule for $W_{:k}$ becomes:

$$W_{:k}^{(t,l+1)} = \max \left( 0, \frac{(XH^T)_{:k} - \sum_{i=0}^{k-1} (HH^T)_{ik} W_{:i}^{(t,l+1)} - \sum_{i=k+1}^{R} (HH^T)_{ik} W_{:i}^{(t,l)} - \alpha_{sp}}{(HH^T)_{kk}} \right) \quad (7)$$

This update rule is computed from the projected gradient rule of the new objective function. The sparsity coefficient $\alpha_{sp}$ was set with performance tests, but was typically low (positive and close to zero) to avoid vanishing columns. We will denote this technique "$l_1$ Penalty Sparse".

- We have proposed a third model, still coming from the idea of keeping the most salient peaks.

To this extent, rather than keeping a fixed number of elements in each column, this algorithm keeps the greatest elements such that the $l_2$ norm of this elements is greater than a constant times the $l_2$ norm of the column. Mathematically, this means that, after computing $W_{:k}$, the final activation column $W_{:k}^{sparse}$ contains the minimal number of the greatest elements of $W_{:k}$ such that $||W_{:k}^{sparse}||_2 \geq \beta_{sp}||W_{:k}||_2$. $\beta_{sp}$ was set with performance tests, but was typically close and lower to 1, in order to keep the most salient peaks. We will denote this technique "$l_2$ Power Sparse".

1st and 3rd model consists both in removing low values after the update. One of the major drawbacks of that technique is that sparsity is enforced on the columns of $H$ while HALS updates iteratively its rows. This implies to add a new loop over the rows of $H$ in the algorithm, losing the conditions of convergence and increasing the complexity of the algorithm.

### 5.1.3 Dictionary learning

In order to enforce the note-frequency codebook factorization, we implemented another approach: rather than finding the frequency-note codebook $W$ as one term of the NMF optimization, we decided to pre-learn it on isolated notes from MAPS [3]. In this technique, rank-one NMF are computed on isolated notes, leading to a frequency column vector codebook, and an activation row vector. The pre-learned $W$ matrix is the result of the concatenation of these frequency column vectors. Finally, the NMF problem is reduced to a nonnegative least square optimization, where only $H$ is updated.

In this model, fundamental frequencies are no longer estimated, but directly retrieved from the annotations of the training set in MAPS, which contains the MIDI value.

We decided to call that technique "Semi-supervised NMF", as the matrix $W$ is no longer estimated but directly computed from MAPS data. It then acts like a dictionary, which should contains all possible MIDI notes and the associated spectrograms.

In our experiments, $W$ is computed from 88 different notes (the entire MIDI scale) of the ENSTDkAm piano in MAPS, which are stereo recordings of a physical piano. More precisely, the notes are the Forte notes of the NO folder, which correspond to notes played without pedal, by contrast with from sustained notes, notes with articulation (staccato for example) or repeated notes. The result is presented in Fig. 8 in a color mesh, where highest values are represented by warm colours.

$H$ was then computed by Nonnegative Least Square using the accelerated HALS update rule, with and without sparsity, with $W$ being fixed.

## 5.2 NMF evaluation summary

To compare these different implementation, we performed tests, which results are presented here. Results represent the average of the measure obtained on the first 30 seconds of each music in the folder "ENSTDkAm" of MAPS. We computed thresholds with different Decibel Reduction Level $\delta_{dB}$ between 10dB and 25dB.

Firstly, we decided to used only one threshold technique on all tests, for consistence between the implementation. By evaluating both thresholds, we concluded that the "fixed" technique seems to perform better on our excerpts, and also seems more stable to intra-excerpt variability than the "adaptive" technique. This conclusion doesn't come from a dedicated test campaign but from little tests over a subset of MAPS.

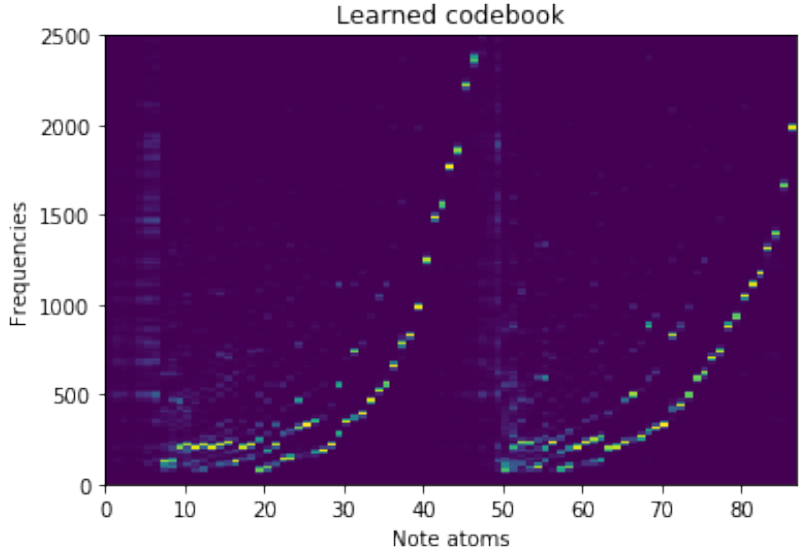

Figure 8: Visualization of the learned codebook

Secondly, we compared the different sparsity constraints, in order to keep only the more promising one in the other tests. We made these tests by learning $W$ and applying sparsity constraints on $H$. This way, we hoped to reduce bias uncorrelated from sparsity. The codebook $W$ was learned on isolated notes recorded in the same conditions than the tests audio excerpts, that is isolated notes from the "ENSTDkAm" folder.

We used the following hyper parameters for each method:

- $n = 20$ for the "$l_0$ Hard Sparse", where sparsity is enforced by keeping the $n$ highest values.

- $\alpha_{sp} = 1e-5$ for the "$l_1$ Penalty Sparse", where sparsity is enforced with the update rule in the Eq. 7.

- $\beta_{sp} = 95$ for the "$l_2$ Power Sparse", where sparsity is enforced by $||W_{:k}^{sparse}||_2 \geq \beta_{sp}||W_{:k}||_2$.

These sparse models were compared to an "unconstrained" model, corresponding to the classic formulation of the optimization problem, corresponding to NNLS optimization on $H$ with the update rule defined in the Eq. 6.

Each of these parameters corresponded to a good trade-off in enforcing sparsity without vanishing the coefficients. These tests gave us the following results:

Table 1: Comparison of sparsity methods, applied on $H$ with $W$ pre-learned.

| Sparsity | $\delta_{dB}$ | Precision | Recall | F-Measure |
|---|---|---|---|---|
| Unconstrained | 17.5 | 0.679 | 0.574 | 0.601 |
| $l_0$ Hard Sparse | 18 | 0.641 | 0.595 | 0.595 |
| $l_1$ Penalty Sparse | 17.5 | 0.702 | 0.569 | 0.609 |
| $l_2$ Power Sparse | 18 | 0.555 | 0.519 | 0.515 |

With these results, we concluded that the $l_1$ penalty method seems more promising than the others and wasn't more computationally expensive than the standard NMF. In addition, this is the only technique preserving the theoretical guarantees of convergence. In that sense, we kept only this method in further work, and will be simplified as "Sparse" in the rest of this report.

Nonetheless, these tests doesn't show a clear benefit of sparsity constraints.

Finally, we compared two NMF techniques:

- "Blind" NMF, where both factors $W$ and $H$ are updated, corresponding to the mathematical model presented above,

- "Semi-supervised NMF", where $W$ is pre-learned on isolated notes, and $H$ the result of NNLS optimization on the music.

To study the bias of learning on a piano, we computed tests by learning $W$ on the same piano than the one used for transcription (ENSTDkAm) and on different pianos (AkPnCGdD and AkPnBcht).

Table 2: Comparison of NMF methods.

| Technique | | $\delta_{dB}$ | Precision | Recall | F-Measure |
|---|---|---|---|---|---|
| Blind NMF | Unconstrained | 13.5 | 0.433 | 0.417 | 0.397 |
| | Only $H$ Sparse | 22.5 | 0.444 | 0.319 | 0.356 |
| | $W$ and $H$ Sparse | 20 | 0.532 | 0.253 | 0.325 |
| Semi-supervised NMF on ENSTDkAm | Unconstrained | 17.5 | 0.679 | 0.574 | 0.601 |
| | $H$ Sparse | 17.5 | 0.702 | 0.569 | 0.609 |
| Semi-supervised NMF on AkPnCGdD | Unconstrained | 16 | 0.478 | 0.49 | 0.454 |
| | $H$ Sparse | 16.5 | 0.471 | 0.489 | 0.454 |
| Semi-supervised NMF on AkPnBcht | Unconstrained | 18.5 | 0.398 | 0.475 | 0.407 |
| | $H$ Sparse | 19.5 | 0.389 | 0.477 | 0.409 |

These results show clear benefits from learning the notes on the piano before transcribing. Conversely, learning on another piano than the one used for transcription seems not to improve the transcription. We can't conclude about the best method between the "Blind" NMF and semi-supervised learning on a different piano than the one to transcribe, as standard deviations of these F measures are around 0.1, which is bigger than the difference between the results.

Though, these results match with the intuition of Sec. 3.3.2, that a preliminary learning of the notes could overfit the spectral shape of a piano. However, it should be noticed that the recording conditions of the samples are hardly comparable. Indeed, the ENSTDkAm samples are recordings of a physical piano, while the AkPnCGdD and AkPnBcht samples are computed from a software. Even if the stereo evaluation (see Sec. 6.1.1) indicates that all samples simulate physical recording conditions, we can't accurately compare such different situations. Still, as ENSTDkAm samples were recorded in a studio, the "studio" recording condition in AkPnCGdD seems a priori closer to ENSTDkAm recording condition than the "Concert Hall" recording condition in AkPnBcht, and could explain the higher results obtained in this configuration.

As learning and transcribing on the same piano outperforms Blind NMF, we decided to keep this configuration for further work.

Nonetheless, these NMF methods are based on mono channel sources, while today, musical recording are mostly multi-channel (principally stereo, which means two channels). In order to exploit the relations between several channels, we have adapted NMF to 3 dimensional models, integrating the channel dimension. This is the sense of the next part of this report.

# 6 Multi channel model

An important part of this work has been the extension and/or development of transcription techniques to multi channel signals. A multi channel signal is a signal containing several tracks or microphone recordings. Generally, the different channels contents are highly correlated, as the result of a mixture of sources between several microphones, or a remastering of the same audio content, particularly in music.

To accurately extend the mono channel model, a prior work is to evaluate the inter channel differences, and then, develop a mathematical model which accurately represents the multi channel information contained in the audio mixture.

## 6.1 Modeling the correlation between channels in stereo signals

This work is focused on the study of stereo signals, which means signals with only two channels. To evaluate the inter channel relationship between both channels, we have to physically model the signals. This part aims at presenting different audio mixtures models.

In all this study, we assume that the recording conditions are time invariant (the sources or the microphones do not physically move during the recording), and that the sources will be only composed of musical sounds, and not external talk or noise.

### 6.1.1 Point source model

In a first approximation, we assume the source to be a point source, meaning that it can assimilated to a point in the space. In that sense, we denote $s(t)$ the source of the signal, and $m_1(t)$, $m_2(t)$ the temporal signal of each channel, as recorded by two microphones. $S(n,f)$, $M_1(n,f)$ and $M_2(n,f)$ represent the STFT of these signals, with $n$ the discrete index of time.

In that representation, the recorded signals are only composed of the source, convoluted with the impulse response of the room at the microphones' position, denoted respectively $v_1$ and $v_2$. These impulse responses are assimilated to filters. Hence, $m_1(t) = [v_1 * s](t)$ and $m_2(t) = [v_2 * s](t)$. In the Fourier domain, the equation become $M_1(n,f) = V_1(n,f)S(n,f)$, and similarly for the second microphone, as the convolution in time domain becomes a product.

As the recording conditions remain the same during the samples, and because we assume that the filter length is smaller than the STFT window size (64ms), assumption called the "narrowband approximation", we assume both $V_i(n,f)$ to be time invariant. Hence, these two assumptions lead to $M_i(n,f) = V_i(f)S(n,f)$.

Finally, let's denote $R_{1/2}(n,f) = \frac{|M_1(n,f)|}{|M_2(n,f)|}$ the inter channel ratio. This inter channel ratio represents the relation between both channel frequency decomposition, and so, notewise, will represent the spectrum differences. Hence, this ratio indicates how to model the frequency differences between both channels. More specifically, we're searching for the most precise way of modeling different codebooks for each channel, so as to fit the spectrums differences. Here, in the point source model:

$$R_{1/2}(n,f) = \frac{|V_1(f)S(n,f)|}{|V_2(f)S(n,f)|} = \frac{|V_1(f)|}{|V_2(f)|}. \tag{8}$$

In this equation, the ratio $R_{1/2}(f)$ is time invariant. Hence, in this model, all inter channel information can be expressed in the frequency domain.

Two types of audio mixes exist with this relation:

- Instantaneous mixture, where the impulse response is a Dirac in time, centered on zero, $\delta(t)$. In the Fourier domain, a Dirac becomes a complex exponential, times a constant. Hence, the magnitude of impulse response in the Fourier domain is the multiplicative constant. This means that the inter channel ratio is constant, so that both channels are proportional. In that case, taking the inter channel information doesn't bring any additional information compared to mono channel model, and is helpless for our task. Instantaneous mixtures generally come from outdoors recordings, or are artificial, like in studio mixtures.

- Convolutive mixture, where the impulse response is a function of the frequency. This is the general case in Eq. 8, and corresponds to more realistic recording conditions, including sound reverberation. In this model, as $R_{1/2}$ only depends on frequency, the spectrums differences can be expressed frequency wise. The instantaneous mixture is an interesting particular case of convolutive mix.

In this model, the inter channel differences can be modeled, at most, by a frequency-wise matrix.

### 6.1.2 Non point sources: notewise point sources model

A more general model is to consider that the source can't be assimilated to a point in the space. In that case, the model can quickly become very complex, particularly time variant, and its exhaustive study would go beyond my internship work. Instead, we focus on a particular case of non point sources by assuming that the source evaluated at a fixed pitch $s_j(t)$ is a point source. In that framework, we rewrite both $m_i(t)$ as $m_1(t) = \sum_{j=1}^{J}[v_{1j} * s_j](t)$ and $m_2(t) = \sum_{j=1}^{J}[v_{2j} * s_j](t)$, with $j$ the pitches index. Hence, evaluating the source in the monophonic case is a point source model as above, and the different channels become $M_i(n, f) = \sum_{j=1}^{J} V_{ij}(f)S(n, f)$ in the Fourier domain. We will called this model "notewise point sources model". This is a frequent model in sound source separation [48]. In order to facilitate notations, we denote $V_i^{(pitch)}(f)$ the filter applied to microphone $i$ at *pitch* fixed. We assume that the function $pitch \mapsto V_i^{(pitch)}$ is not constant, as this case is analogous to the precedent model.

Notewise, the audio mix can either be instantaneous or convolutive.

- If the mixture is instantaneous, it means that $f \mapsto V_i^{(pitch)}(f)$ is constant, and so that inter channel differences can be modeled by a notewise constant in each channel, and so, model the codebook differences with an additional notewise matrix. We denote this model "notewise instantaneous model".

- If the mixture is convolutive and not instantaneous, these inter channel differences would be pitch and frequency variant. In that sense, modeling a codebook for each channel can describe these differences. We denote this model "notewise convolutive model".

This notewise point sources model considers that inter channel relations can be modeled considering only the pitch of the played note and the frequency decomposition of the signal in both channels. In that sense, this model is particularly adapted for NMF-like models, as all the inter channel information can be contained in frequency and channel wise matrices.

### 6.1.3 Notewise convolutive time-varying model

In addition of these time invariant models, we considered a time variant model, based on Neuroimaging data [49]. The idea is to study the inter channel ratio $f \mapsto R_{1/2}^{(\theta, pitch)}(f)$ at fixed *pitch*, and to consider the time coordinate $\theta$ as as delay from onset instead of an absolute time. If, at different absolute time coordinates, but equally spaced $\theta$ from the onset of a same *pitch*, the ratio has the same frequency shape, then these relations could be modeled by a delay-pitch convolutive filter. This is convenient as this convolutive filter would be fixed notewise, which is our framework.

### 6.1.4 Other models

If any regular shape or note wise shape can't be computed, we will considered the underlying multi channel model as too complex. Indeed, if any particular shape can be derived from the note, a same note will have different shapes depending on the moment it is played in at least one of the channels. This would mean, in other words, that a same note could not be factorized by a frequency decomposition, making NMF underdetermined.

These theoretical considerations conclude that depending on the type of stereo of the excerpt, the multi-channel model should be chosen carefully, as it could be, for the better, non-optimized, and for the worst, counterproductive (too constrained for example).

## 6.2 Case of MAPS signals

In order to develop the adequate model for MAPS, we have to know what is the type of stereo and the information added when considering both channels. As in our previous study, the recording conditions are time invariant (the source and the microphone do not move with time), and the excerpts are recorded in a closed room [3].

Firstly, we computed the ratio $R_{1/2}(t, f) = \frac{|M_1(t,f)|}{|M_2(t,f)|}$ on isolated notes, and compare it on several successive time index. An example of this computation is presented in the Fig. 9.

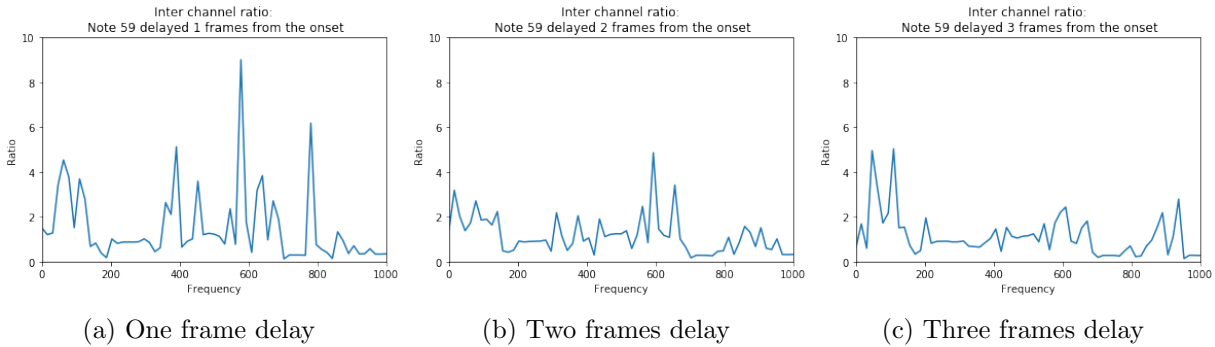

(a) One frame delay  (b) Two frames delay  (c) Three frames delay

Figure 9: Computation of the Inter channel ratio on the note 59 from the folder "ISOL/NO" of MAPS, with different delays after the onset

We conclude, with this study, that this ratio is time variant. Hence, the notewise point sources model presented above can't be applied in perfect match. Still, the underlying models will be tested, as comparison with mono channel models.

Secondly, we compared different ratios of the same note at the same delay from the onset, on notes from the "RE" folder of MAPS. This tests aimed at finding regular shapes of the notes spectrums, as discussed in Sec. 6.1.1. A result is presented in Fig. 10.

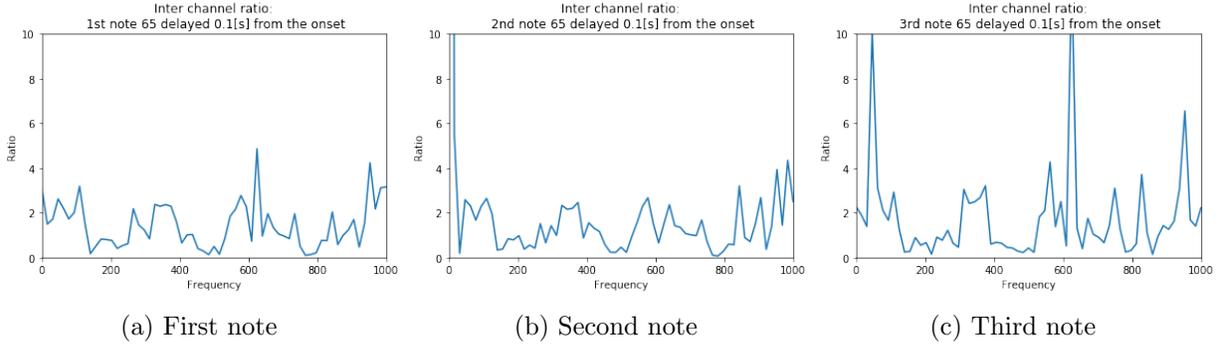

(a) First note      (b) Second note      (c) Third note

Figure 10: Computation of the inter channel ratio on the note 65 from the folder "ISOL/RE" of MAPS, with a delay of 0.1s after the onset. The note are the first three appearing in the sample.

We concluded, with this study, that the notewise inter channel relations seems to have regular shapes with time, or at least similar enough to be considered as a notewise convolutive time-varying model. It is hard to be strongly affirmative though, as the shapes aren't exactly the same.

In that sense, in order to effectively compute the relations between the different channels in MAPS, the most adapted multi-channel model seems to be be the model from Morup & al. [49].

Starting from the theoretical considerations, we will present several multi-channel models in the following parts. Particularly, we focused our efforts on models adapted to the notewise point sources model. We will then test them on MAPS excerpts, even if the audio mixture doesn't seem to be the most adapted.

## 6.3 Extending NMF for multi-channel transcription

A natural way of adding the channel dimension in NMF would be the computation of different NMF on each channel, and to combine the results. As our goal is to exploit the inter channel information, we didn't considered that this technique could significantly improve our results, and decided not to implement it.

A more constraining model, that we will call "Simultaneous NMF", is proposed by Alves for drum transcription [50]. This model considers all channels at once for the decomposition, in a unique NMF model. This is obtained by decomposing a spectrogram $\tilde{X} = \begin{bmatrix} X_1 \\ X_2 \\ \ldots \\ X_C \end{bmatrix}$, concatenation of independent STFT on each of the $C$ channels, in $\tilde{W}H$, with $\tilde{W} = \begin{bmatrix} W_1 \\ W_2 \\ \ldots \\ W_C \end{bmatrix}$, frequency decomposition of the notes in each channel, and $H$, time activation of these atoms, as in the precedent NMF model. More precisely, in our stereo model, this results in:

$$\begin{bmatrix} X_1 \\ X_2 \end{bmatrix} = \begin{bmatrix} W_1 \\ W_2 \end{bmatrix} H. \tag{9}$$

In that technique, the $W_i$ are not learned independently and stacked, but learned at the same time, on spectrograms constituted of both channels. This way, by normalizing the spectrograms before concatenating them, the loudness differences in both channels are reduced.

This model represents the notewise convolutive model.

Nonetheless, this model assumes that notes are played at the same time in both channels, as $H$ is shared between channels. This assumption can be made for the samples in MAPS, as the channels correspond to two microphones, recording the same piano, and at approximately the same distance, or at least close enough to make the assumption that the delay will be smaller than the time window.

In this paper, this technique is compared to another model, more specific to drum transcription. In drums recording in general, a microphone is fixed on every drum. In that sense, the author makes the assumption that these microphones contain exactly one drum (or a drum extremely dominant). This way, it can computes the transcription by thresholding only these microphones. This so-called "naive" multichannel method performs relatively good, but is outperformed by its Simultaneous NMF. In addition, it is not relevant in piano transcription.

Implemented and solved as our previous NMF model, see Sec. 5.1.1, this model performs equivalently:

Table 3: Simultaneous NMF results on MAPS.

| Simultaneous NMF | $\delta_{dB}$ | Precision | Recall | F-Measure |
|---|---|---|---|---|
| Unconstrained | 17 | 0.678 | 0.570 | 0.598 |
| $H$ Sparse | 17.5 | 0.665 | 0.581 | 0.6 |

These results can be explained by the fact that the stereo model in MAPS is not adapted to this model, as concluded in Sec. 6.2, but that it allows a great variability between channels, in order to correctly fit the respective decompositions.

## 6.4 Tensor model

By considering the channel dimension, the model becomes 3 dimensional (time/frequency/channel), with a collection of spectrograms according to the channels. In that sense, instead of crafting matrices by concatenating two dimensions as in Sec. 6.3, we propose to extend the model to the study of 3 dimensional objects.

Tensors are mathematical objects, defined as "multidimensional arrays" [51]. A n-th order tensor is defined as an element of the space given by the tensor product of n vector spaces. As so, a first-order tensor is a vector, and a second-order a matrix. For our task, we are interested in third-order tensors, and I will only study such tensors in this report.

### 6.4.1 Tensor decomposition

By definition, a third-order tensor is an array defined with three indices. It can also be seen as a "concatenation" of matrices, the same way that a matrix is a concatenation of vectors. Fig. 11 provides visualizations of a third-order tensor.

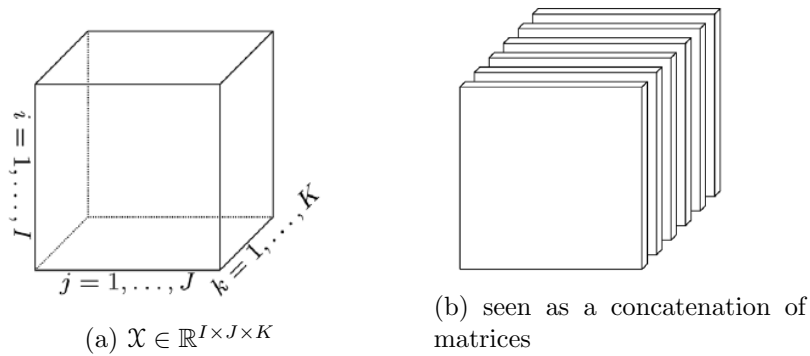

(a) $\mathcal{X} \in \mathbb{R}^{I \times J \times K}$

(b) seen as a concatenation of matrices

Figure 11: A third order tensor (from [51])

The outer product of two vectors is the product resulting in a matrix. The outer product of two vectors $u$ and $v$ is the matrix $W$ where each element $w_{ij} = u_i v_j$. The outer product is denoted as $W = u \circ v$. Similarly, the outer product of three vectors gives a third-order tensor. The converse is not true: most tensors cannot be written as the outer product of three vectors.

Similarly to rank-1 matrices, where all column vectors are linearly dependant (and so can be written as the outer product of two vectors), a rank-1 third-order tensor is defined as a third-order tensor which can be written as an outer product of three vectors. Fig. 12 shows a rank-1 tensor decomposition.

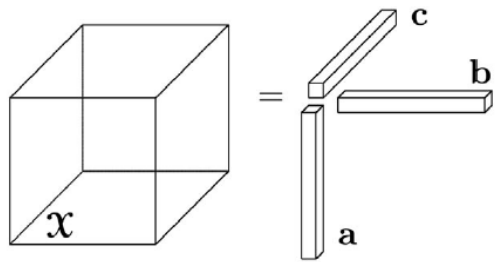

Figure 12: A rank 1 third-order tensor $\mathcal{X}$ (from [51]). The $(i, j, k)$ element $x_{ijk}$ of $\mathcal{X}$ is equal to $a_i b_j c_k$

Intuitively, rank-1 tensors seem easier to study. In that sense was proposed the idea of decomposing any tensor as a finite sum of rank-one tensors. This decomposition holds many names [51]: the polyadic form of a tensor, CANDECOMP for canonical decomposition [52] or also PARAFAC for parallel factors [53]. It is even sometimes called CP decomposition for CANDECOMP/PARAFAC decomposition as the two names were introduced in the same year (1970) [54]. Finally, the concept remains the same. For sake of simplicity, I will keep the name of polyadic form or polyadic decomposition as in [55], because it was the first name given to this idea. It is represented in Fig. 13.

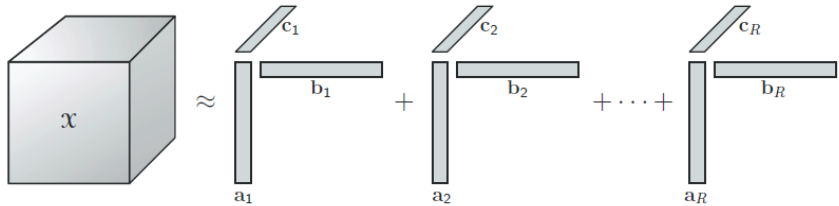

Figure 13: A polyadic decomposition of a tensor $\mathcal{X}$ (from [51])

In Fig. 13, the decomposition is an approximation. In fact, with a rank $R$ large enough, any tensor $\mathcal{X}$ can be decomposed in an "exact" polyadic form, meaning that the equation in the decomposition is an equality, i.e for any $\mathcal{X} \in \mathbb{R}^{I \times J \times K}$, $\mathcal{X} = \sum_{r=1}^{R} a_r \circ b_r \circ c_r$ with $a_r \in \mathbb{R}^I, b_r \in \mathbb{R}^J, c_r \in \mathbb{R}^K$, or elementwise, $x_{ijk} = \sum_{r=1}^{R} a_{ir} b_{jr} c_{kr}$. Hereby, the smallest number of rank-1 tensors for the decomposition to be exact defines the rank of the tensor, noted $rank(\mathcal{X})$. Such a decomposition is called the rank decomposition. The decomposition is often written $[A, B, C]$, with $A, B$ and $C$ respectively the matrices $[a_1, ..., a_R], [b_1, ..., b_R], [c_1, ..., c_R]$. A convenient writing from [55] is, given a tensor $\mathcal{X} \in \mathbb{R}^{I \times J \times K}, \mathcal{X} = (A \otimes B \otimes C) \mathcal{I}_R$ with $A \in \mathbb{R}^{I \times R}, B \in \mathbb{R}^{K \times R}, C \in \mathbb{R}^{K \times R}$ and $\mathcal{I}_R$ the third-order diagonal tensor of ones, which is $\sum_{r=1}^{R} e_r \otimes e_r \otimes e_r$. The product $.\otimes.$ is an operator tensor product, derived from the outer product $.\circ.$, and supply this convenient notation. It is defined in [56].

Rank decomposition offers interesting results, the most interesting one being that it is often unique. Unique means here that two different solutions of the problem can exist, but that they only differ by permutations on $A, B$ and $C$ (as long as the same permutation is applied to the three of them) and/or by a rescaling of the different matrices, as $\sum_{r=1}^{R} (\alpha_r a_r) \circ (\beta_r b_r) \circ (\gamma_r c_r)$ with $\alpha_r \beta_r \gamma_r = 1$ for every $r$. "Often" unique means here that the uniqueness conditions are much weaker for tensors than for matrices. Without developing them, sufficient or necessary conditions only come in general from the rank of the matrices $A$, $B$ and $C$, whereas the same results for matrices has to be obtained with constraints over the values of the decomposition matrices (for example constraints of orthogonality).

This is the main interest for the use of tensors, because global minima seem easier to reach than in the NMF. Nonetheless, it is proven that finding the rank of a tensor is NP-hard [57]. Still, it is proven that $rank(\mathcal{X}) \leq \min(IJ, JK, IK)$, for any $\mathcal{X} \in \mathbb{R}^{I \times J \times K}$, bounding the search.

Because of the complexity of the problem, there is no algorithm for determining the rank of a tensor. Hence, most algorithms fit polyadic decomposition with different number of components, and stop when one is found to be good enough. Indeed, even if a perfect decomposition exists in theory, the noise in data in practice makes it counter-productive to find it.

When the number of rank-1 tensors is fixed, some algorithms are efficient in finding a good decomposition. In [51], the Alternating Least Square algorithm (ALS) is presented as the most used algorithm. This algorithm searches for the $\tilde{\mathcal{X}} = \sum_{r=1}^{R} \lambda_r a_r \circ b_r \circ c_r$ minimizing $\|\mathcal{X} - \tilde{\mathcal{X}}\|_F^2 = \sum_{i,j,k}(\mathcal{X} - ABC)_{ijk}^2$. ALS alternatively fixes two of the three matrices and solve the problem for the last one, then iterates over the three matrices and finally loop until a stop condition is reached, as presented for the matrix case in the Alg. 1.

As tensor decomposition provides interesting properties compared to matrix decomposition, notably milder uniqueness conditions, we decided to evaluate it on multi-channel transcription. In order to study more precisely tensors, notably developing optimization algorithms, some notions of tensor algebra have to be introduced.

### 6.4.2 Notions of tensor algebra

Tensors, as multi dimensional arrays, need an algebra to be studied. The goal of this part is to present some useful operations to study the optimisation models using tensors, and not developing the entire theoretical model. This algebra excerpt is extended from matrices, and largely refer to it. Actually, an important operation consists in reordering the elements of a tensor in order to transform it in a matrix. This operation is called "matricization", or unfolding [51]. This can be viewed as a flattening of all modes of the tensor, except one: given a tensor of size $(I_0, I_1, \cdots, I_N)$, the n-mode unfolding of this tensor will be a matrix of size $(I_n, I_0 \times I_1 \times \cdots \times I_{n-1} \times I_{n+1} \cdots \times I_N)$. In our third-order model, the three n-mode unfolding of the tensor $\mathcal{X} \in \mathbb{R}^{I \times J \times K}$ are matrices of respective size $(I, J \times K)$, $(J, I \times K)$ and $(K, I \times J)$. As an example, let's consider a tensor $\mathcal{T}$ of size $(3, 4, 2)$, which both slides are respectively:

$$T_1 = \begin{bmatrix} 0 & 2 & 4 & 6 \\ 8 & 10 & 12 & 14 \\ 16 & 18 & 20 & 22 \end{bmatrix} \text{ and } T_2 = \begin{bmatrix} 1 & 3 & 5 & 7 \\ 9 & 11 & 13 & 15 \\ 17 & 19 & 21 & 23 \end{bmatrix}.$$

The 1-mode unfolding will be the matrix $T_{(1)}$, of size $(3, 8)$:

$$\mathcal{T}_{(1)} = \begin{bmatrix} 0 & 1 & 2 & 3 & 4 & 5 & 6 & 7 \\ 8 & 9 & 10 & 11 & 12 & 13 & 14 & 15 \\ 16 & 17 & 18 & 19 & 20 & 21 & 22 & 23 \end{bmatrix} \tag{10}$$

Analogously, the 2-mode and 3-mode unfolding of $\mathcal{T}$ will be respectively the matrices $T_{(2)}$, of size $(4, 6)$:

$$\mathcal{T}_{(2)} = \begin{bmatrix} 0 & 1 & 8 & 9 & 16 & 17 \\ 2 & 3 & 10 & 11 & 18 & 19 \\ 4 & 5 & 12 & 13 & 20 & 21 \\ 6 & 7 & 14 & 15 & 22 & 23 \end{bmatrix}$$

and $T_{(3)}$, of size $(2, 12)$:

$$\mathcal{T}_{(3)} = \begin{bmatrix} 0 & 2 & 4 & 6 & 8 & 10 & 12 & 14 & 16 & 18 & 20 & 22 \\ 1 & 3 & 5 & 7 & 9 & 11 & 13 & 15 & 17 & 19 & 21 & 23 \end{bmatrix}.$$

As long as notations and equations are consistent, the order of unfolding is not important. In fact, different notations exist in litterature, as in [51], which presents unfolding with a different set of permutations of columns.

This specific order of n-mode unfolding, presented above, is defined in the Tensorly toolbox documentation [58] and by Cohen [56] as mapping the element at index $(i_0, i_1, \cdots, i_N)$ to the element at index $(i_n, j)$, with $j = \sum_{k=0, k \neq n}^{N} i_k \times \prod_{m=k+1, k \neq n}^{N} I_m$. As the order of unfolding impacts the operating spaces, it impacts the products and properties. The products adapted to this order of unfolding are defined in [56].

Unfolded, tensors can be studied as matrices, and so evaluated with usual products, that we will reintroduce.

Firstly, the Kronecker product: given two matrices $A \in \mathbb{R}_+^{I \times J}$ and $B \in \mathbb{R}_+^{K \times L}$, the Kronecker product is denoted $A \boxtimes B \in \mathbb{R}_+^{(IK) \times (JL)}$. It is defined as:

$$A \boxtimes B = \begin{bmatrix} a_{11}B & a_{12}B & \cdots & a_{1J}B \\ a_{21}B & a_{22}B & \cdots & a_{2J}B \\ \vdots & \vdots & \ddots & \vdots \\ a_{I1}B & a_{I2}B & \cdots & a_{IJ}B \end{bmatrix}, \tag{11}$$

or, by its explicit form:

$$A \boxtimes B = \begin{bmatrix} a_{11}b_{11} & a_{11}b_{12} & \cdots & a_{11}b_{1L} & \cdots & \cdots & a_{1J}b_{11} & a_{1J}b_{12} & \cdots & a_{1J}b_{1L} \\ a_{11}b_{21} & a_{11}b_{22} & \cdots & a_{11}b_{2L} & \cdots & \cdots & a_{1J}b_{21} & a_{1J}b_{22} & \cdots & a_{1J}b_{2L} \\ \vdots & \vdots & \ddots & \vdots & & & \vdots & \vdots & \ddots & \vdots \\ a_{11}b_{K1} & a_{11}b_{K2} & \cdots & a_{11}b_{KL} & \cdots & \cdots & a_{1J}b_{K1} & a_{1J}b_{K2} & \cdots & a_{1J}b_{KL} \\ \vdots & \vdots & & \vdots & \ddots & & \vdots & \vdots & & \vdots \\ \vdots & \vdots & & \vdots & & \ddots & \vdots & \vdots & & \vdots \\ a_{I1}b_{11} & a_{I1}b_{12} & \cdots & a_{I1}b_{1L} & \cdots & \cdots & a_{IJ}b_{11} & a_{IJ}b_{12} & \cdots & a_{IJ}b_{1L} \\ a_{I1}b_{21} & a_{I1}b_{22} & \cdots & a_{I1}b_{2L} & \cdots & \cdots & a_{IJ}b_{21} & a_{IJ}b_{22} & \cdots & a_{IJ}b_{2L} \\ \vdots & \vdots & \ddots & \vdots & & & \vdots & \vdots & \ddots & \vdots \\ a_{I1}b_{K1} & a_{I1}b_{K2} & \cdots & a_{I1}b_{KL} & \cdots & \cdots & a_{IJ}b_{K1} & a_{IJ}b_{K2} & \cdots & a_{IJ}b_{KL} \end{bmatrix} \tag{12}$$

We define the Khatri-Rao product as the columnwise Kronecker product of two matrices. Given two matrices $A \in \mathbb{R}_+^{I \times K}$ and $B \in \mathbb{R}_+^{J \times K}$, with the same number of columns, the Khatri-Rao product, denoted $A \odot B \in \mathbb{R}_+^{(IJ) \times K}$, is defined as:

$$A \odot B = \begin{bmatrix} A_{:,1} \boxtimes B_{:,1} & A_{:,2} \boxtimes B_{:,2} & \cdots & A_{:,K} \boxtimes B_{:,K} \end{bmatrix} \tag{13}$$

or, in a more explicit form:

$$A \odot B = \begin{bmatrix} a_{11}b_{11} & a_{12}b_{12} & \cdots & a_{1K}b_{1K} \\ a_{11}b_{21} & a_{12}b_{22} & \cdots & a_{1K}b_{2K} \\ \vdots & \vdots & \ddots & \vdots \\ a_{11}b_{J1} & a_{12}b_{J2} & \cdots & a_{1K}b_{JK} \\ \vdots & \vdots & & \vdots \\ \vdots & \vdots & & \vdots \\ a_{I1}b_{11} & a_{I2}b_{12} & \cdots & a_{IK}b_{1K} \\ a_{I1}b_{21} & a_{I2}b_{22} & \cdots & a_{IK}b_{2K} \\ \vdots & \vdots & \ddots & \vdots \\ a_{I1}b_{J1} & a_{I2}b_{J2} & \cdots & a_{IK}b_{JK} \end{bmatrix} \tag{14}$$

Assuming that a tensor $\mathcal{X}$ can be decomposed in an exact polyadic form of three matrices $A,B$ and $C$ such as $\mathcal{X} = (A \otimes B \otimes C)\mathcal{I}_R$, we can deduce from these definitions that:

$$\mathcal{X}_{(1)} = A(B \odot C)^T. \tag{15}$$

Analogously:

$$\mathcal{X}_{(2)} = B(A \odot C)^T \text{ and } \mathcal{X}_{(3)} = C(A \odot B)^T.$$

Analogously to the unfolding of vectors, we call "vectorization" the operation of transforming a tensor into a column vector by stacking its columns [56]. This operation is denoted $vec(.)$.

Particularly, given three matrices $A$, $B$ and $X$, coupling vectorization and the Kronecker product leads to the following equation [56]:

$$vec(AXB) = (A \boxtimes B^T)vec(X) \tag{16}$$

Particularly, if $X$ is diagonal:

$$vec(AXB) = (A \odot B^T)diag(X) \tag{17}$$

The proof of this equation is given in Appendix B.

With these theoretical considerations, we have implemented some tensor models to the transcription problem. One particular model is the NTF, standing for Nonnegative Tensor Factorization, which is the direct analogy of NMF in tensor algebra. This model corresponds to the polyadic form of a tensor with nonnegativity constraints.

### 6.4.3 NTF

Extending the NMF model in third-order tensor algebra, the spectrogram matrix becomes a spectrogram tensor $\mathcal{X}$, composed of the spectrograms of each channel along the third mode, as represented in Fig. 14.

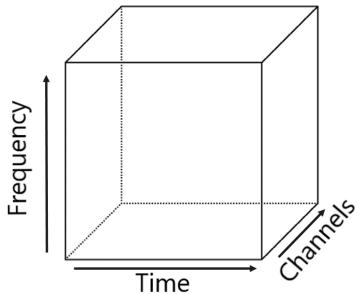

Figure 14: Representation of the spectrogram tensor $\mathcal{X}$

With the same notations than in Sec. 3.3.1, the NTF model is expressed: $\mathcal{X} \approx (W \otimes H \otimes Q)\mathcal{I}_R$ with $W \in \mathbb{R}_+^{K \times R}$, $H \in \mathbb{R}_+^{N \times R}$ and $Q \in \mathbb{R}_+^{C \times R}$, nonnegative element-wise [48]. In the same context than in Sec. 3.3.2, it can be seen as the dictionary decomposition and activations of different channels in music, $Q$ representing the activation level in each channels. Hence, the polyadic decomposition of such a tensor can be written as $\mathcal{X} \approx \sum_{r=1}^{R} w_r \circ h_r \circ q_r$ with $w_r, h_r, q_r$ respectively the frequency decomposition, the activation in time and the weight in each channel of the $r$ note.

With these notations, computing NTF can be cast as the following minimization problem:

$$\min_{W \geq 0, H \geq 0, Q \geq 0} \|\mathcal{X} - (W \otimes H \otimes Q)\mathcal{I}_R\|_F^2,$$

with $\mathcal{I}_R$ the third-order diagonal tensor of ones.

In my internship, we integrated the Tensorly toolbox to perform tensor operations [58], and, to solve the NTF problem, we decided to adapt the Hierarchical Alternative Least Squares algorithm (Alg. 2) used for NMF. Indeed, with the Eq. 15, we obtain:

$$\|\mathcal{X} - (W \otimes H \otimes Q)\mathcal{I}_R\|_F^2 = \|\mathcal{X}_{(1)} - W(H \odot Q)^T\|_F^2, \qquad (18)$$

and analogously for the two other modes. In other words, the tensor model has been rewritten in a matrix form, analogous to the standard NMF problem. As presented in [51], we can then solve this optimization problem by adapting the NNLS algorithm and by fixing two factors when updating the third one. This results in Alg. 4. It should be noted that [51] uses a different unfolding order, leading to a different order of the factors in the Khatri-Rao product.

---

**Algorithm 4** NNLS adaptation for NTF

---

**Input:** Nonnegative tensor $\mathcal{X} \in \mathbb{R}^{K \times N \times C}$ and a factorization rank $R$
**Output:** $W \in \mathbb{R}_+^{K \times R}, H \in \mathbb{R}_+^{N \times R}, Q \in \mathbb{R}_+^{C \times R}$, such that $\mathcal{X} \approx (W \otimes H \otimes Q)\mathcal{I}_R$
1: Generate $W^{(0)}$, $H^{(0)}$ and $Q^{(0)}$, initial matrices.
2: **for** $t = 1, 2, ...$ **do**
3: $\quad W^{(t)} = update\left(\mathcal{X}_{(1)}, (H^{(t-1)} \odot Q^{(t-1)})^T, W^{(t-1)}\right)$
4: $\quad H^{(t)} = update\left(\mathcal{X}_{(2)}, (W^{(t)} \odot Q^{(t-1)})^T, H^{(t-1)}\right)$
5: $\quad Q^{(t)} = update\left(\mathcal{X}_{(3)}, (W^{(t)} \odot H^{(t)})^T, Q^{(t-1)}\right)$

---

The update rule used in my internship is still the Accelerated HALS, as presented for $W$ in the Alg. 3.

Table 4: NTF results on MAPS.

| NTF | $\delta_{dB}$ | Precision | Recall | F-Measure |
|---|---|---|---|---|
| Unconstrained | 17 | 0.626 | 0.537 | 0.555 |
| $H$ Sparse | 17 | 0.644 | 0.536 | 0.563 |

In this model, the inter channel relations are modeled by a notewise constant. Hence, this model is adapted for notewise instantaneous mixtures, which corresponds to multichannel signals where the inter channel ratio is time and frequency invariant. While time invariance is a common assumption in audio signals, see Sec. 6.1.1, frequency invariance seems inadapted and over-constraining the model, leading to poor results. In that sense, we decided to test another model, more relaxed about the frequency invariance constraint, called PARAFAC2 [59].

### 6.4.4 PARAFAC2

PARAFAC2 is a relaxed formulation of NTF, where one factor admits small variations in the different modes [59]. In that model, instead of factorizing a tensor $\mathcal{T}$ as $\mathcal{T}_{ijk} \approx \sum_{r=1}^{R} a_{ir} b_{jr} c_{kr}$ as in NTF, this tensor is factorized $\mathcal{T}_{ijk} \approx \sum_{r=1}^{R} a_{ir}^k b_{jr} c_{kr}$ in PARAFAC2, with $A^k$ varying according to the mode $k$. In this model though, the different $A^k$ matrices are coupled by a common latent matrix $A^*$, with $A_{:l_1}^{k}{}^T A_{:l_2}^{k} = A_{:l_1}^{*}{}^T A_{:l_2}^{*}$, or, elementwise, $\sum_{m=1}^{M} a_{ml_1}^k a_{ml_2}^k = \sum_{r=1}^{R} a_{rl_1}^* a_{rl_2}^*$, for every index $l_1$, $l_2$ and $k$, and denoting $M$ and $R$ respectively the number of lines of $A^k$ and $A^*$. The latent matrix $A^*$ is a square matrix of size $R \times R$.

In our transcription model, we implemented PARAFAC2 with the coupling over the codebooks, to approach the fit of a notewise convolutive model in each channel, while keeping both matrices coupled. In that sense, with our notation, PARAFAC2 model is written:

$$\mathcal{X}_{fnk} \approx \sum_{r=1}^{R} w_{fr}^k h_{rn} q_{kr}$$

$$\text{with } \sum_{f=1}^{F} w_{fp_1}^k w_{fp_2}^k = \sum_{r=1}^{R} w_{rp_1}^* w_{rp_2}^* \quad \forall p_1, p_2 \in [\![1, \cdots R]\!], \forall k \in [\![1, \cdots C]\!] \tag{19}$$

Instead of solving this tensor formulation of the model, we will again convert it in a matrix formulation. To correctly formulate this inter channel variation of the codebook, we rewrite the model channelwise. Hence, denoting $X_k$ the $k$-th slice of the tensor (spectrogram of the $k$-th channel), it becomes $X_k \approx W_k D_k H$, with $W_k = P_k W^*$ and $P_k^T P_k = I_R$. As $W_k$ is of size $F \times R$ and $W^*$ of size $R \times R$, $P_k$ is of size $F \times R$. $D_k$ is a diagonal matrix of size $R \times R$ containing on its diagonal the coefficients of the $k$-th line of $Q$, or, in other terms, the weight of each note in the channel $k$.

Hence, the objective function to minimize for computing PARAFAC2 is $\sum_{k=1}^{C} ||X_k - P_k W^* D_k H||_F^2$ subject to $P_k^T P_k = I_R, \quad \forall k \in [\![1, \cdots C]\!]$.

Imposing the coupling of all $W_k$ matrices is a hard constraint, and may be too hard as the $P_k$ need to be orthogonal, which impose for the columns of $W_k$ to be transformed similarly. In that sense, we implemented a relaxed version of the problem, called Flexible PARAFAC2 [60]. Instead of imposing the coupling, this method only imposes convergence towards this coupling by penalizing the objective function, which becomes

$$\sum_{k=1}^{C} ||X_k - W_k D_k H||_F^2 + \mu_k ||W_k - P_k W^*||_F^2 \tag{20}$$

with:

$$\begin{aligned} X_k &\approx W_k D_k H \\ W_k &\approx P_k W^* \quad \text{with } P_k^T P_k = I_R, \quad \forall k \in [\![1, \cdots C]\!] \end{aligned} \tag{21}$$

Parameter $\mu_k$ is the penalty factor controlling priority towards fitting the spectrogram or coupling of matrices. The bigger, the most important the coupling is. In that sense, $\mu_k$ is generally increased with the iteration, as the reconstruction error decreases. In the process, matrices $W_k$, $D_k$ are normalized columnwise for the norm to influence the activations in $H$, and influence the detection of notes.

**Algorithm 5** NNLS adaptation for PARAFAC2
___
**Input:** Nonnegative tensor $\mathcal{X} \in \mathbb{R}^{K \times N \times C}$ and a factorization rank $R$
**Output:** $W_k \in \mathbb{R}_+^{K \times R}$, $D_k \in \mathbb{R}_+^{R \times R}$, $H \in \mathbb{R}_+^{N \times R}$, such that $X_k \approx W_k D_k H$, and $P_k$, $W^*$ such that $W_k \approx P_k W^*$ for every $k$.
1: Generate $P_k^{(0)}$, $W_k^{(0)}$, $D_k^{(0)}$, and $H^{(0)}$, initial matrices.
2: **for** $t = 1, 2, ...$ **do**
3: $\quad W^* = \frac{1}{\sum_{k=1}^C \mu_k} \sum_{k=1}^C \mu_k P_k^T W_k$
4: $\quad$ **for** $k = 1, 2, \cdots, C$ **do**
5: $\quad\quad P_k^{(t)} = U_{:,1:R} V^T{}_{1:R,:}$ with $(U, S, V^T) = \text{SVD}\left(W_k^{(t-1)} W^{*T}\right)$
6: $\quad\quad W_k^{(t)} = \underset{W_k^{(t-1)} \geq 0}{\arg\min} \left\|X_k - W_k^{(t-1)} D_k^{(t-1)} H^{(t-1)}\right\|_F^2 + \mu_k \left\|W_k^{(t-1)} - P_k^{(t)} W^*\right\|_F^2$
7: $\quad\quad\quad\quad\quad\quad\quad\quad\quad\quad\quad\quad\quad\quad\quad\quad$ ▷ This is solved with a special HALS, see Alg. 6.
8: $\quad\quad D_k^{(t)} = update\left(vec(X_k), W_k^{(t)} \odot H^{(t-1)^T}, diag(D_k)\right)$
9: $\quad\quad\quad\quad\quad\quad\quad\quad\quad\quad\quad\quad\quad\quad\quad\quad$ ▷ The accelerated HALS *update*, see Alg. 3
10: $\quad H^{(t)} = \underset{H^{(t-1)} \geq 0}{\arg\min} \sum_{k=1}^C \|X_k - W_k^{(t)} D_k^{(t)} H^{(t-1)}\|_F^2$
11: $\quad\quad\quad\quad\quad\quad\quad\quad\quad\quad\quad\quad\quad\quad\quad\quad$ ▷ H updates simultaneously on every $k$.
___

This problem is solved by updating the factors successively, as detailed in Alg. 5.

The update rule of the diagonal factor $D_k$ is obtained by developing the generic update formulation, $D_k = \underset{D_k \geq 0}{\arg\min} \|X_k - W_k D_k H\|_F^2$. Obviously, vectorizing the matrices doesn't alter the equation, and so $\|X_k - W_k D_k H\|_F^2 = \|vec(X_k) - vec(W_k D_k H)\|_2^2$. Using Prop. 17, as $D_k$ is a diagonal matrix, $vec(W_k D_k H) = (W_k \odot H^T) diag(D_k)$. Finally, the update rule of $D_k$ can be written as:

$$D_k = \underset{D_k \geq 0}{\arg\min} \|vec(X_k) - (W_k \odot H^T) diag(D_k)\|_2^2, \tag{22}$$

which can be computed similarly to previous algorithms.

As $W_k$ is updated following a different objective function than the usual one, the HALS algorithm has to be adapted. In that sense, by computing the projected gradient on this objective function: $\left\|X_k - W_k^{(t-1)} D_k^{(t-1)} H^{(t-1)}\right\|_F^2 + \mu_k \left\|W_k^{(t-1)} - P_k^{(t)} W^*\right\|_F^2$, we obtain the Alg. 6.

Finally, $H$ is updated by HALS, as in Alg. 2, by computing $C = \sum_{k=1}^C W_k^{(t+1)^T} X_k$ and $D = \sum_{k=1}^C W_k^{(t+1)^T} W_k^{(t+1)}$

Tested on the first 30 second of all MAPS, this model performs worst than our other multi-channel models.

In fact, these results can be explained by the fact that updating the learned codebooks will

**Algorithm 6** Updating $W_k$

**Input:** Nonnegative matrix $X_k \in \mathbb{R}^{K \times N}$, $W_k^{(t-1)}$, $\mu_k$, $P_k$, $W^*$, and a factorization rank $R$
**Output:** $W_k^{(t)} \in \mathbb{R}_+^{K \times R}$.

1: Compute $A = X_k H^{(t)T}$, $B = H^{(t)} H^{(t)T}$ and $M = P_k W^*$.
2: **for** $i = 1, 2, ..., R$ **do**
3:     **if** $B_{ii} \neq 0$ **then**
4:         $W_{k:i}^{(t+1)} = \max\left(0, \dfrac{A_{:i} - \sum_{j=0}^{i-1} B_{ji} W_{k:j}^{(t+1)} - \sum_{j=i+1}^{R} B_{ji} W_{k:j}^{(t)} + \mu_k M_{:i}}{B_{ii} + \mu_k}\right)$

Table 5: Flexible PARAFAC2 results on MAPS.

| Flexible PARAFAC2 | $\delta_{dB}$ | Precision | Recall | F-Measure |
|---|---|---|---|---|
| Unconstrained | 17 | 0.601 | 0.531 | 0.539 |
| $H$ Sparse | 16.5 | 0.547 | 0.458 | 0.478 |

only deteriorate their fit of the note spectrums decomposition. In that sense, PARAFAC2 benefits should be greater in an unsupervised case.

# 7 Perspective

Beyond this internship, further work can be completed on multi-channel adaptation of NMF for music transcription. Notably, we will evoke some avenues that follow the work presented in this report.

## 7.1 Transcription

First, we observed in this work the importance of the note detection technique on the results. Indeed, our thresholding technique was very sensitive to little variations, and needs feedback to be correctly set. This is prohibitive for developing an unsupervised transcription method. In that sense, a different thresholding technique must be found, or combined with a normalisation phase which stretches the gap between activation and noise.

In this problematic too, different note detecting techniques seem promising. For example, [61] proposed a framework with derivatives of the activation for event detection and a learning phase on isolated notes computing an adapted threshold; or also the attack/decay model proposed in [38], which decompose the note in two parts in order to approach the physical note representation, which should perform better on event detection.

These two perspectives should directly impact our transcription results, but shouldn't change the conclusions on the comparison between the different models.

## 7.2 Multi-channel tensor models

A first perspective should be the validation of these tensor models on the specific kind of audio mixtures they should fit. Indeed, by testing on MAPS uniquely, with its specific type of audio mixture, benefits of multi-channel models could not be confirmed.

In addition, the multi channel transcription task could be extended by developing other tensor models.

Notably, a work could be made on relaxing the learning phase. Dictionary learning improvements on statistical outputs are remarkable in this context, but learning the entire spectral decomposition may be an overconstraining solution to the transcription problem, as it can be hard to access to all notes in both channels. It can also overfit the data, as changing the recording condition or the piano can strongly affect the transcription quality, as the semi-supervised NMF on different pianos showed.

In that sense, an interesting implementation could be made by extending PARAFAC2 to exploit the regular shape of a note and its harmonics in both channels. Indeed, rather than learning the entire codebooks, we could use the harmonic shape of each note to find accurate codebooks on the data. Notably, using the latent matrix to represent note harmonic combs, like in [26], could enhance physically relevant note decomposition in the codebooks, while the optimization problem will fit the spectral shape of each channel.

Finally, an interesting model, as said in Sec. 6.2, could be the convolutional CP [49], which seems promising to better fit audio mixtures such as MAPS.

# 8 Conclusion

In conclusion, Music Transcription face numerous problems in the case of polyphonic sounds, as the mix of the partials of different notes generally fools the automatic recognition system, leading to numerous transcription errors. The current systems provide insufficient results for the problem to be solved. In that sense, my internship focused on implementing and testing NMF with sparsity constraints and notes spectrum decomposition learning in a first part, and developing multi-channel transcription models in a second part. The conducted experimentations show a clear benefit from dictionary learning among the other techniques. Conversely, as dictionary learning simplifies the problem, the multi-channel models presented in this report do not improve transcription quality. Still, it could improve classical NMF models in the case of complex audio mixtures, as it could represent time varying frequency decompositions, which isn't possible in the NMF framework. In addition, the different multi-channel models are adapted to represent multi-channel signals of different natures, and, as a more constraining model, could improve blind decomposition, whether in music transcription or in another domain.

## A  Concatenation of the results of all techniques

Table 6: Comparison of all tested methods.

| Technique | | $\delta_{dB}$ | Precision | Recall | F-Measure |
|---|---|---|---|---|---|
| Blind NMF | Unconstrained | 13.5 | 0.433 | 0.417 | 0.397 |
| | Only $H$ Sparse | 22.5 | 0.444 | 0.319 | 0.356 |
| | $W$ and $H$ Sparse | 20 | 0.532 | 0.253 | 0.325 |
| Semi-supervised NMF on ENSTDkAm | Unconstrained | 17.5 | 0.679 | 0.574 | 0.601 |
| | $H$ Sparse | 17.5 | 0.702 | 0.569 | 0.609 |
| Semi-supervised NMF on AkPnCGdD | Unconstrained | 16 | 0.478 | 0.490 | 0.454 |
| | $H$ Sparse | 16.5 | 0.471 | 0.489 | 0.454 |
| Semi-supervised NMF on AkPnBcht | Unconstrained | 18.5 | 0.398 | 0.475 | 0.407 |
| | $H$ Sparse | 19.5 | 0.389 | 0.477 | 0.409 |
| Simultaneous NMF | Unconstrained | 17 | 0.678 | 0.570 | 0.598 |
| | $H$ Sparse | 17.5 | 0.665 | 0.581 | 0.6 |
| NTF | Unconstrained | 17 | 0.626 | 0.537 | 0.555 |
| | $H$ Sparse | 17 | 0.644 | 0.536 | 0.563 |
| PARAFAC2 | Unconstrained | 17 | 0.601 | 0.531 | 0.539 |
| | $H$ Sparse | 16.5 | 0.547 | 0.458 | 0.478 |

# B  Proof of Property 17

*Démonstration.*

$$(A \boxtimes B^T)vec(X) = \begin{bmatrix} a_{11}b_{11} & a_{11}b_{21} & \cdots & a_{11}b_{K1} & \cdots & \cdots & a_{1K}b_{11} & a_{1K}b_{21} & \cdots & a_{1K}b_{K1} \\ a_{11}b_{12} & a_{11}b_{22} & \cdots & a_{11}b_{K2} & \cdots & \cdots & a_{1K}b_{12} & a_{1K}b_{22} & \cdots & a_{1K}b_{K2} \\ \vdots & \vdots & \ddots & \vdots & & & \vdots & \vdots & \ddots & \vdots \\ a_{11}b_{1J} & a_{11}b_{2J} & \cdots & a_{11}b_{KJ} & \cdots & \cdots & a_{1K}b_{1J} & a_{1K}b_{2J} & \cdots & a_{1K}b_{KJ} \\ \vdots & \vdots & & \vdots & \ddots & & \vdots & \vdots & & \vdots \\ \vdots & \vdots & & \vdots & & \ddots & \vdots & \vdots & & \vdots \\ a_{I1}b_{11} & a_{I1}b_{21} & \cdots & a_{I1}b_{K1} & \cdots & \cdots & a_{IK}b_{11} & a_{IK}b_{21} & \cdots & a_{IK}b_{K1} \\ a_{I1}b_{12} & a_{I1}b_{22} & \cdots & a_{I1}b_{K2} & \cdots & \cdots & a_{IK}b_{12} & a_{IK}b_{22} & \cdots & a_{IK}b_{K2} \\ \vdots & \vdots & \ddots & \vdots & & & \vdots & \vdots & \ddots & \vdots \\ a_{I1}b_{1J} & a_{I1}b_{2J} & \cdots & a_{I1}b_{KJ} & \cdots & \cdots & a_{IK}b_{1J} & a_{IK}b_{2J} & \cdots & a_{IK}b_{KJ} \end{bmatrix} \begin{bmatrix} x_1 \\ 0 \\ \vdots \\ 0 \\ 0 \\ x_2 \\ 0 \\ \vdots \\ \vdots \\ 0 \\ x_K \end{bmatrix}$$

$$= \begin{bmatrix} a_{11}b_{11}x_1 + a_{12}b_{21}x_2 + \cdots + a_{1K}b_{K1}x_K \\ a_{11}b_{12}x_1 + a_{12}b_{22}x_2 + \cdots + a_{1K}b_{K2}x_K \\ \vdots \\ a_{11}b_{1J}x_1 + a_{12}b_{2J}x_2 + \cdots + a_{1K}b_{KJ}x_K \\ \vdots \\ \vdots \\ a_{I1}b_{11}x_1 + a_{I2}b_{21}x_2 + \cdots + a_{IK}b_{K1}x_K \\ a_{I1}b_{1J}x_1 + a_{I2}b_{2J}x_2 + \cdots + a_{IK}b_{KJ}x_K \end{bmatrix}$$

$$= \begin{bmatrix} a_{11}b_{11} & a_{12}b_{21} & \cdots & a_{1K}b_{K1} \\ a_{11}b_{12} & a_{12}b_{22} & \cdots & a_{1K}b_{K2} \\ \vdots & \vdots & \ddots & \vdots \\ a_{11}b_{1J} & a_{12}b_{2J} & \cdots & a_{1K}b_{KJ} \\ \vdots & \vdots & & \vdots \\ \vdots & \vdots & & \vdots \\ a_{I1}b_{11} & a_{I2}b_{21} & \cdots & a_{IK}b_{K1} \\ a_{I1}b_{1J} & a_{I2}b_{2J} & \cdots & a_{IK}b_{KJ} \end{bmatrix} \begin{bmatrix} x_1 \\ x_2 \\ \vdots \\ x_K \end{bmatrix} = (A \odot B^T)diag(X)$$

□